\documentclass[acmsmall]{acmart}
\AtBeginDocument{%
  \providecommand\BibTeX{{%
    \normalfont B\kern-0.5em{\scshape i\kern-0.25em b}\kern-0.8em\TeX}}}

\setcopyright{acmcopyright}
\copyrightyear{2022}
\acmYear{2022}
\acmDOI{10.1145/1122445.1122456}

\acmJournal{TOSEM} 
\acmVolume{37}
\acmNumber{4}
\acmArticle{111}
\acmMonth{11}

\usepackage{microtype}
\usepackage{textcomp}
\usepackage{bm}
\usepackage[caption=false]{subfig}
\usepackage{flushend}
\usepackage{csquotes}
\usepackage{booktabs} 
\usepackage{framed}
\usepackage{graphicx}
\usepackage{tabularx}
\usepackage{longtable}
\usepackage{graphicx}
\usepackage{url}
\usepackage{multirow}
\usepackage{enumitem}
\usepackage{hyperref}
\usepackage{lscape}
\usepackage[normalem]{ulem}
\useunder{\uline}{\ul}{}
\usepackage{latexsym}
\usepackage{xcolor}
\definecolor{lightgray}{gray}{0.9}
 
\usepackage{siunitx}  
\usepackage{textgreek}  
\usepackage{wasysym}
\usepackage{makecell}

\usepackage{tabularx}

\begin{document}

\title{A Theory of Scrum Team Effectiveness}

\author{Christiaan Verwijs}
\email{christiaan.verwijs@theliberators.com}
\affiliation{%
  \institution{The Liberators}
  \country{The Netherlands}
}

\author{Daniel Russo}
\authornote{Corresponding author.}
\email{daniel.russo@cs.aau.dk}
\orcid{0000-0001-7253-101X}
\affiliation{%
  \institution{Department of Computer Science, Aalborg University}
  \streetaddress{A.C. Meyers Vaenge, 15, 2450}
  \city{Copenhagen}
  \country{Denmark}}

\renewcommand{\shortauthors}{Verwijs and Russo, 2022}

\begin{abstract}
Scrum teams are at the heart of the Scrum framework. Nevertheless, an integrated and systemic theory that can explain what makes some Scrum teams more effective than others is still missing. 
To address this gap, we performed a seven-year-long mixed-method investigation composed of two main phases. First, we induced a theoretical model from thirteen exploratory field studies. Our model proposes that the effectiveness of Scrum teams depends on five high-level factors - responsiveness, stakeholder concern, continuous improvement, team autonomy, and management support - and thirteen lower-level factors.
In the second phase of our study, we validated our model with a Covariance-Based Structural Equation Modeling (SEM) analysis using data from about 5,000 developers and 2,000 Scrum teams that we gathered with a custom-built survey. 
Results suggest a very good fit of the empirical data in our theoretical model ($CFI = 0.959, RMSEA = 0.038, SRMR = 0.035$).
Accordingly, this research allowed us to (1) propose and validate a generalizable theory for effective Scrum teams and (2) formulate clear recommendations for how organizations can better support Scrum teams.
\end{abstract}



\maketitle

\section{Introduction}
\label{sec:Introduction}

Software development teams are at the core of the Scrum framework~\cite{schwaber1995}. A recent investigation by Russo~\cite{russo2021PLS} identified Scrum teams as the most important driver of project success~\cite{russo2021ASM}.
Like other Agile development methods, Scrum follows a collaborative, iterative, and human-oriented approach to software development by reflecting the principles of the Agile Manifesto~\cite{AgileManifesto2001}. 

Among Agile methods, Scrum is the most popular~\cite{version1stateofAgile}. 
According to market research led by VersionOne, in 2020, 76\% of organizations reported that they use Scrum or a hybrid version of it. Software houses routinely use it for their internal development process~\cite{williams2011scrum}.
Scrum teams focus on delivering incremental business value through the continuous interaction with stakeholders within an established framework~\cite{schwaber1995}. With a minimal set of prescribed roles and events, it is up to teams to build on this foundation by adding practices for stakeholder collaboration, estimation, task breakdown, portfolio management, and so on that make sense in their context.
Thus, to be effective, Scrum assumes a high degree of autonomy for teams to perform their work~\cite{hoda2017becoming}.
The 2020 Scrum Guide highlights the importance of Scrum teams over the individual roles that make it up~\cite{schwaber2020scrum}. 
By shifting control from middle management to autonomous Scrum teams, the role of management has become less relevant.
Therefore, the contemporary understanding of Scrum is to support teams to deliver and deploy software as frequently as is helpful~\cite{hoda2017becoming}, which is summarized in the 2020 Scrum Guide as ``Sprints are the heartbeat by which ideas are turned into value.''~\cite{schwaber2020scrum}.

In the last twenty years, researchers performed several empirical investigations into Scrum, mostly field studies~\cite{edison2021comparing}. 
So far, research typically focused on process improvement factors (such as security~\cite{pohl2015b}), transformation processes~\cite{dikert2016challenges}, methodological and tailoring practices~\cite{messina2016new}, or Scrum artifacts~\cite{ciancarini2018Agile}.
Some scholars also explored teaming aspects, like autonomy~\cite{moe2010teamwork} and proactive behavior~\cite{junker2021Agile}.
However, few studies have specifically explored which team-level factors are most important to Scrum team effectiveness, with few exceptions~\cite{moe2010teamwork,moe2008understanding}.
Additionally, most works performed so far are not generalizable because they are field studies or use a research strategy not aimed at generalizing results~\cite{stol2018abc,edison2021comparing}. 
Thus, we do not know if the accumulated knowledge about Scrum teams is specific to a few contexts or if it is generalizable to larger populations.
A sample study, using adequate statistical techniques to test research hypotheses~\cite{russo2019soft}, would bridge this gap~\cite{stol2018abc}.
Accordingly, we formulate our Research Questions (RQ):

\begin{itemize}
\item RQ$_1$: \textit{Which are the key factors that determine Scrum team effectiveness, and how do they relate to each other?}
\item RQ$_2$: \textit{How generalizable is the theoretical model for Scrum teams?}
\end{itemize}

This study focuses on Scrum teams, and not more broadly on Agile teams of other kinds, for two reasons. The first is that Scrum provides a well-defined framework with a small set of roles and events for teams to implement. This brings focus and a shared vocabulary to the investigation. The second reason is that the prevalence of Scrum makes it easier to reach a sufficiently large sample.

The characteristic of sample studies is that they rely on pre-existing theoretical evidence~\cite{russo2021PLS}.
Accordingly, we had to induce a working theory first to address RQ$_1$.
To do so, we performed 13 case studies from 2014 to 2019, involving informants from over 45 Scrum teams. 
Using the multiple cases study approach by Eisenhardt~\cite{eisenhardt2007theory}, we inferred a theoretical model from our qualitative findings and extant literature to explain how team-level factors contribute to Scrum team effectiveness.

In this model, we define \textit{Scrum team effectiveness} as the satisfaction of team members with their work process and stakeholders' satisfaction with the outcomes of that process. We found that team effectiveness is determined by five high-level factors; the ability of a team to be responsive to changing needs of stakeholders, the concern that teams have for the needs of their stakeholders, a climate of continuous improvement, team autonomy, and management support.

To answer RQ$_2$, we ran a large-scale cross-sectional study by collecting data from 4,940 software professionals of 1,978 Scrum teams, resulting in one of the most extensive sample studies performed in software engineering. We aimed to test whether the inferred theoretical model also fit empirical data and matched the day-to-day reality of Scrum teams. Using Covariance Based Structural Equation Modeling (CB-SEM or SEM in short), we found that the model fit the data well and that its proposed factors explained a substantial amount of variance in Scrum team effectiveness.
A replication package is also openly available on Zenodo to encourage secondary studies.


In the remainder of this paper, we describe the related work in Section~~\ref{sec:related}.
Afterward, we introduce our Mixed-Methods research design in Section~~\ref{sec:researchdesign} and induce our theoretical model from 13 multiple exploratory case studies in Section~~\ref{sec:multiplecasestudy}.
Subsequently, in Section~~\ref{sec:SEM}, we validate our model through a large-scale cross-sectional study using Covariance Based Structural Equation Modeling.
Finally, we discuss the implications for research and practice along with the study limitations in Section~~\ref{sec:Discussion} and draw our conclusion by outlying future research directions in Section~~\ref{sec:Conclusion}.

\section{Related work}
\label{sec:related}

Software development teams have been at the core of Scrum since the first guide was issued in 2010 by Jeff Sutherland and Ken Schwaber\footnote{www.scrumguides.org/revisions.html}.
Consequently, software engineering literature investigated several aspects of Scrum teams over the last decade.
We searched for peer-reviewed publications in Scopus to find previous work, where we found 24 relevant papers. 
Since this investigation seeks relevance instead of completeness, we primarily address studies that focus on Scrum teams to identify our research gap.

Agile software development in general, and Scrum in particular, are widely studied in software engineering research~\cite{edison2021comparing}.
Much of this work has focused on the transition from plan-driven to Agile approaches at the organizational level. Dikert, Paasivaara \& Lassenius~\cite{dikert2016challenges} and López-Martínez et al. ~\cite{lopez2016problems} identified common challenges in Agile transformations, such as lack of management support, limited understanding of Agile methodologies, and a cultural mismatch with Agile principles. However, plan-driven development may co-exist with Agile practices if management takes proper actions to mitigate challenges and confusion that might arise in Scrum teams~\cite{van2013Agile}. Another challenge is that Agile practices do not guide HR processes such as onboarding and knowledge transfer, so organizations have to figure out how to deal with it~\cite{donmez2013practice}. Paasivara \& Lassenius~\cite{paasivaara2016scaling} investigated challenges related to inter-team coordination when scaling Scrum across many teams. Other studies have investigated requirements engineering in Agile environments~\cite{schon2017Agile} and how Agile methodologies impact contracting~\cite{russo2018contracting}. Recently, the Agile Success Model was proposed by Russo~\cite{russo2021ASM} to provide a general understanding of how the dynamics within an Agile project lead to its success. This study identified (Scrum) teams and the skills of their members as the most critical enabler.

Scholars have also investigated the transition from plan-driven to Agile approaches at the team level. Hoda \& Noble~\cite{hoda2017becoming"R} used a Grounded Theory approach to understand how Agile teams transition from traditional to Agile practices. Gren et al.~\cite{gren2017group} suggest that group developmental aspects of Scrum teams are critical success factors during the transition from a plan-driven to an Agile development fashion.

Effective self-managing teams are at the center of the Scrum framework and other Agile methodologies. Surprisingly few studies have attempted to conceptualize Scrum team effectiveness and its determinants, even though such insights are invaluable to practitioners. Although productivity-based measured such as the SPACE framework~\cite{forsgren2021space} have been proposed, they are difficult to aggregate and compare across organizations and are often constrained by factors beyond a team's control~\cite{mathieu2008team}. Melo et al. \cite{melo2013interpretative} proposed a more comprehensive definition of productivity of Agile teams by including outcomes such as customer satisfaction, creativity, and product quality.

Moe et al. ~\cite{moe2008understanding,moe2010teamwork} were among the first to investigate Scrum team effectiveness from the perspective of \textit{teamwork}. Salas et al. ~\cite{salas2005there} define this as ``the set of interrelated thoughts, actions, and feelings of each team member that are needed to function as a team and that combine to facilitate coordinated, adaptive performance and task objectives resulting in value-added outcomes''. They propose a model for effective teamwork based on five core components (shared leadership, peer feedback, redundancy, adaptability, and team orientation) and three coordinating mechanisms (mutual trust, shared mental models, and closed-loop communication). This model has been successfully applied to Scrum teams by Moe \& Dings{\o}yr ~\cite{moe2008understanding} and Agile teams by Strode, Dings{\o}yr \& Lindsorn~\cite{strode2022teamwork}.

A strength of this perspective on Scrum team effectiveness is that it grounds it in a tradition of teamwork research. However, the generalized model for effective teamwork by Salas et al. ~\cite{salas2005there} may not consider idiosyncrasies of Scrum teams that are critical to their effectiveness.
We consider this particularly relevant to three areas.
First, it does not address how teams interact with their stakeholders. Several studies have emphasized the role of the Product Owner and shared Product Ownership as a solid contributor to team effectiveness~\cite{bass2018empirical,kristinsdottir2016responsibilities,rolland2016tailoring}. This is particularly relevant to Scrum teams as their interactions with stakeholders are a critical part of Agile methodologies~\cite{AgileManifesto2001}.
Second, the capability of Scrum teams to be responsive to changing needs in their environments is not considered, even though this is an important reason to use Agile methodologies over plan-driven approaches~\cite{donmez2013practice}. 
Third, Scrum teams are always embedded in larger systems that constrain or enable their effectiveness~\cite{decuyper2010grasping}. For example, scholars such as Argyris~\cite{argyris1999organizational} argue that continuous improvement is difficult without a learning culture~\cite{watkins1996action,garvin2008yours}. Similarly, the self-managing nature of Scrum teams demands a different approach from management in order for teams to make their own decisions. Manz et al. ~\cite{manz1987leading} describe this shift as ``leading others to lead''.

These areas are illustrative of a systems perspective on teams~\cite{luhmann1995social}. Although effective teamwork within a team is undoubtedly a solid contributor to team effectiveness, much of it will be negotiated on the boundaries with the larger systems it is part of~\cite{decuyper2010grasping}. For example, what support do teams receive from the organization? How do they collaborate with stakeholders, customers, and users? Furthermore, how is their effectiveness enabled or impeded by technology and organizational culture?

Thus, this study aims to provide a systemic understanding of the factors that contribute to the effectiveness of Scrum teams. This includes factors internal to teams and those within the boundaries of teams.

\section{Research Design}
\label{sec:researchdesign}
\begin{figure*}
\centering
\includegraphics[height=1.5in]{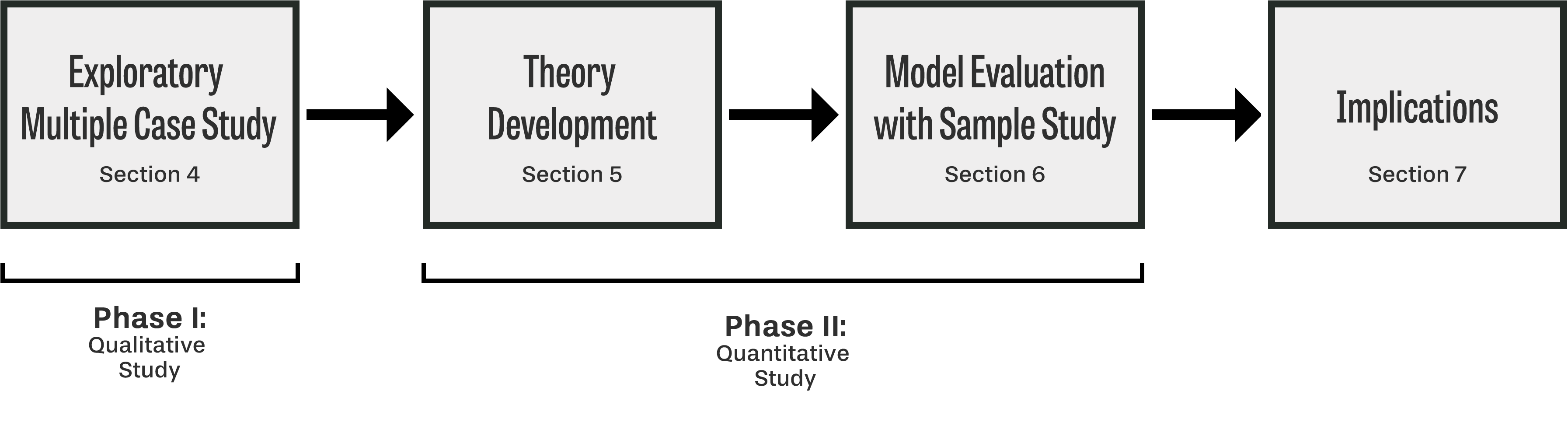}
\caption{The design of this two-phased study, and the corresponding sections in this paper.}
\label{fig:steps}
\end{figure*}

We deployed a Mixed Method approach to answer our research questions~\cite{Creswell2007}. Such approaches are increasingly used in the social sciences, humanities~\cite{johnson2007toward}, and also in software engineering~\cite{stol2017competition,Russo2018ISQ} to increase the fidelity of theories and findings over purely quantitative or qualitative inquiries~\cite{mingers2001combining}. 
The different research phases of our research design are summarized in Figure~\ref{fig:steps}.

In the \textbf{first phase}, we used a multiple case study to identify potentially relevant factors by observing Scrum teams in 13 different cases. We combined our observations with extant literature to induce a preliminary, testable theory (the objective of RQ$_1$). While the natural research setting offered by case studies is excellent for developing such a theoretical understanding of realistic phenomena~\cite{stol2018abc}, it is not aimed at the generalization of findings across settings (the objective of RQ$_2$). Furthermore, the exploratory nature of our multiple case studies would not be sufficient to claim a robust theory. 

In the \textbf{second phase}, we tested the preliminary theoretical model in a neutral research setting~\cite{stol2018abc} using Structural Equation Modeling\footnote{Structural Equation Modeling divide itself into Covariance-Based Structural Equation Modeling and Partial Least Squares Structural Equation Modeling. This paper focuses solely on Covariance-Based Structural Equation Modeling. For an overview of these different computation methods and their terminology, cf. Russo \& Stol, 2021.} (SEM) and a substantial sample of Scrum teams from different settings. Structural Equation Modeling is commonly used to test hypothesized relationships between unobserved (latent) constructs through observed indicator variables~\cite{kline2015principles, byrne2010structural}. The hypothesized relationships in the model are specified into a series of structural (regression) equations that predict the variance-covariance matrix that should be observed between variables in the data if the model fits perfectly. A second variance-covariance matrix is calculated from sample data and compared with the first matrix. The smaller the discrepancy between the matrices, as indicated by the statistical goodness of fit indices~\cite{byrne2010structural}, the better the theoretical model ``fits'' the observed data. A strength of Structural Equation Modeling is that it is an inherently confirmatory, hypothesis-driven approach to model testing~\cite{byrne2010structural}. Because the entire model is tested simultaneously, researchers can perform factor analysis and regression analysis in a single model. Finally, Structural Equation Modeling allows researchers to assess and adjust for measurement error by explicitly estimating it as part of the model, which is impossible in traditional multivariate approaches. This allows researchers to arrive at more accurate estimates, especially when there is a substantial error in the measurements, such as in sample studies~\cite{byrne2010structural}. 

\section{Phase I: Multiple Case Study}

\subsection{The Scrum Framework}
\label{sec:ScrumFramework}
Because this study aims to understand what makes Scrum teams effective, we begin with an overview of the Scrum framework to provide consistent terminology throughout this paper. 

The Scrum framework is designed around self-managing Scrum teams~\cite{schwaber2020scrum}. A \textit{``Scrum team''} consists of a Scrum Master, a Product Owner, and developers. \textit{``Stakeholders''} are external parties with a stake in the outcomes of the team, like customers and users. The Scrum framework prescribes five recurring events to foster collaboration and coordination within the team and its stakeholders. The first recurring event is the \textit{``Sprint''} itself. It has a consistent duration of a month or less. Each Sprint begins with \textit{``Sprint Planning''}, where a team sets a goal on what it aims to deliver that Sprint (the \textit{``Sprint Goal''}) and selects the work to achieve it. This work is selected from a prioritized list called the \textit{``Product Backlog''}. It contains all requirements, features, and ideas deemed valuable for the product. The selection of work for the current Sprint is called the \textit{``Sprint Backlog''}. It changes to reflect discoveries that teams make as they collaborate to achieve the goal. One daily opportunity for such discoveries is the \textit{``Daily Scrum''} where members coordinate their work intending to deliver an increment that achieves their Sprint Goal. These increments need to be done, meaning that they conform to the quality standards set out by the team as captured in their \textit{``Definition of Done''}.
At least before the end of the Sprint, the team shares its increment with stakeholders and gathers feedback during the \textit{``Sprint Review''}. The Product Backlog is updated accordingly to reflect discoveries. Finally, a \textit{``Sprint Retrospective''} is held within the team to learn and improve its process. Within this framework, the \textit{``Scrum Master''} is accountable for enacting Scrum while the \textit{``Product Owner} is accountable that what a team works on is indeed valuable. All other members of the team are considered \textit{``Developers''} of the product and include all the skills and specializations needed to deliver done increments.

\subsection{Case selection}
\label{sec:multiplecasestudy}
This research investigates Scrum team effectiveness.
Consequently, our unit of analysis is the Scrum team.
Due to the exploratory nature of this first phase, we did not use \textit{a priori} criteria to guide our sample selection. 
Instead, we opted for a flexible sampling design and selected cases to maximize case variability and improve our theoretical understanding~\cite{yin1994case}.
In particular, we looked for a variation in size of organizations, commercial or non-commercial nature of the organizations, industry, and age of establishment.
From a theory-building perspective, we relied on Eisenhardt's replication logic~~\cite{eisenhardt1989building} i.e., each experience serves as a different experiment that stands on its own as an analytic unit, inducing theory through observations or data~~\cite{eisenhardt2007theory}.

The sample for our multiple case study consisted of 13 separate cases that took place between 2014 and 2019 (see table~\ref{table:cases}). 
Each case study observation took place for several months to a year. Each case involved one or more Scrum teams in organizations in improving their effectiveness. Some of these organizations already worked with Scrum (Table~\ref{table:cases}: \#10, \#11, \#12), while others implemented it during the case studies.

\begin{table*}
\centering
\small
\caption{Description of the cases used in this study for observations and interviews}
\begin{tabular}{@{}p{0.02\linewidth}p{0.08\linewidth}p{0.25\linewidth}p{0.10\linewidth}p{0.10\linewidth}p{0.30\linewidth}@{}}
\toprule
\textbf{\#} & \textbf{Year} & \textbf{Domain} & \textbf{Org size} & \textbf{Nr. of teams} & \textbf{Roles involved in observations}\\
\midrule 
1 & 2014-2016 & Transportation & 4,000-5,000 & 5 & 2x Scrum Master, 1x Product Owner, 30+ Developers, 1x Program Manager\\ 
2 & 2014-2017 & Software development & 10-20 & 2 & 2x Scrum Masters, 1x Product Owner, 8x Developers, 1x CxO\\ 
3 & 2014-2018 & Software development & 10-20 & 1 & 2x Scrum Masters, 1x Product Owner, 1x CEO\\ 
4 & 2014-2016 & Software development & 10-20 & 1 & 1x Scrum Master, 1x Product Owner, 12x Developers, 1x CxO\\ 
5 & 2015-2017 & Retail & 50-100 & 2-5 & 5x Scrum Masters, 2x Product Owners, 25x Developers, 2x CxO\\ 
6 & 2016-2017 & Information Technology & 100-200 & 5 & 5x Scrum Masters, 4x Product Owners, 25x Developers, 2x CxO\\ 
7 & 2016-2017 & Non-Profit & 10-50 & 1 & 1x Scrum Master, 1x Product Owner, 6x Developers\\ 
8 & 2016-2017 & Healthcare & 50-100 & 2 & 1x Scrum Master, 1x Product Owner, 8x Developers\\ 
9 & 2017-2018 & Utilities & 500 & 1 & 1x Scrum Master, 1x Product Owner, 6x Developers\\ 
10 & 2018-2019 & Tourism industry & 4,000-5,000 & 5 & 12x Scrum Masters\newline \\ 
11 & 2018-2019 & Recruitment & 50-100 & 1 & 1x Scrum Master, 1x Product Owner, 5x Developers, 1x CxO\\ 
12 & 2019 & Retailer & $>10,000$ & 12 & 15x Scrum Masters, 4x Product Owners, 3x Agile Coaches\\ 
13 & 2019 & Research \& education & 1,000-2,000 & 5 & 4x Scrum Masters, 2x Product Owners, 38x Developers, 1x Department Manager\\ 
\bottomrule
\end{tabular}
\label{table:cases}
\end{table*}

\subsection{Collection of Qualitative Data}
We used different techniques and multiple data sources for each case. The use of multiple data sources allows the triangulation of findings and increases internal validity~\cite{stake1995art}. 

Each case began with semi-structured \textbf{interviews} of the informants that led internal change initiatives to increase team effectiveness, usually Scrum Masters, Product Owners, and management. Most interviews lasted one hour. Informants were asked to define effectiveness and identify which factors they thought were relevant. We opted for semi-structured interviews to allow the broadest possible scope for data collection. We periodically interviewed these informants again during our engagements and at the end. After each interview, we summarized key findings and validated them with the informants. In total, we collected 155 pages of field notes and interviews.

We then hosted \textbf{workshops} that were attended by the Scrum teams and relevant stakeholders. The workshops lasted between 1 and 2 days. During the workshops, we moderated a conversation in the group to identify and build consensus on what made them effective and where and how they wanted to improve their process to become more effective. The groups generated \textbf{documentation} in the form of flip charts to capture insights in their own words. Sixty-two documents were created in this manner. We were also given access to other sources of documentation in most cases, such as a Definition of Done, Product Backlogs, developed Increments, and team charters.

Following the initial workshops, we were on-premise for one or more days per week, ranging between two months and a year. The duration depended on the length of the engagement and the organization's size. In most cases, a desk in the team room was provided to us, or a desk in a room close to the teams. This allowed us to capture \textbf{observations} during more than 150 meetings and Scrum Events, as well as social interactions that happened naturally throughout the day (e.g., lunches, breaks, conversations). These observations allowed us to capture relevant group dynamics and intimate project knowledge from the participants and the organization. We captured the observations in field notes directly after they occurred. For case \#12, we could not remain on-premise as this company was geographically distant. Instead, we visited the organization on two occasions and performed follow-up interviews remotely. We included this case due to the unique characteristics of its organizational culture and because the workshops were attended by a substantial cross-section of many Scrum teams and organizational layers.

\subsection{Analysis of Qualitative Data}
We used the analytical framework known as \textit{Gioia methodology}~~\cite{gioia2013seeking} to analyze our qualitative data.
In particular, Gioia et al. proposed a structured approach to foster scientific rigor in qualitative data analysis.
This approach, which is more and more common within the management community, is now also being used in software engineering~\cite{russo2021ASM}.

Following the Gioia methodology, we analyzed our data in three different stages.
In the first stage, we used open coding to identify 61 distinct low-level concepts in the data~\cite{corbin1990grounded}. 
We then used axial coding in the second stage to group related concepts into 16 second-order themes relevant to answering RQ$_1$.
Finally, we grouped second-order themes into six high-level aggregate dimensions until theoretical saturation was achieved. In all stages, we aimed to retain the terminology used by practitioners to keep our theoretical model grounded.
Researchers ideally perform the first and second stages with as little theoretical knowledge as possible to prevent confirmation bias~\cite{gioia2013seeking}. 
In addition, we involved a domain expert unfamiliar with the study to categorize first-order concepts into themes and aggregate dimensions independently. 
Several iterations were performed until consensus was achieved.
This three-stage approach is a data structure where grounded observations about success factors are induced into six high-level theoretical aggregate dimensions connected to existing literature.

\subsection{Results from the case studies}
The following sections report the six aggregate dimensions that we induced from the multiple case studies. Guided by our first research question (RQ$_1$), we begin our examination with team effectiveness and then explore relevant team-level dimensions that may contribute to it. We use the reporting structure recommended by Gioia~~\cite{gioia2013seeking} to provide clarity on how we aggregated our observations into higher-order concepts. We provide observations and exemplary quotes from informants for each concept and connect them to extant literature. The complete data structure is available in Table~\ref{tab:datastructure} in the Appendix.

\subsubsection{Team Effectiveness}
The Scrum teams broadly identified two ultimate outcomes of effective teamwork with Scrum. We summarize the observations and second-order themes in Table~\ref{tab:datastructureteameffectiveness}.

\textbf{Stakeholder satisfaction:}
Most Scrum teams expected that effective teamwork with Scrum results in more satisfied stakeholders, particularly customers and users (see Tab~\ref{table:cases}: \#1, \#2, \#3, \#4, \#5, \#7, \#8, \#12). Many teams reasoned that this was because Scrum allowed them to release increments sooner and more frequently (\#2, \#3, \#4) and deliver value to stakeholders more frequently (Tab~\ref{table:cases}: \#2, \#3, \#4, \#6, \#7). The increased collaboration between teams and stakeholders was another reason. (Tab~\ref{table:cases}: \#2, \#5, \#6).\\
\begin{displayquote}
``A big opportunity [of Scrum] for us is to get more satisfied customers'' [PO - 2]\\
\end{displayquote}
This expected outcome is congruent with the goal of the Scrum Framework, which is to deliver value to stakeholders more frequently~\cite{schwaber2020scrum}. Other studies also identified stakeholder satisfaction as a key indicator for Agile Teams of post-release quality~\cite{kupiainen2014industrial} and team performance~\cite{mahnic2007using}. 

\textbf{Team Morale:}
Another outcome that Scrum teams expected was increased morale and motivation (Tab~\ref{table:cases}: \#2, \#3, \#4, \#5, \#6, \#7, \#8, \#9, \#13). For many Scrum teams, this resulted from increased collaboration (Tab~\ref{table:cases}: \#1, \#3, \#6, \#8). Other reasons that were given ranged from ``do work that matters'' (Tab~\ref{table:cases}: \#6), ``more personal autonomy'' (Tab~\ref{table:cases}: \#2, \#6) to ``getting things done'' (Tab~\ref{table:cases}: \#4, \#6).\\
\begin{displayquote}
``I expect that when we work well with Scrum, our team spirit will be higher'' [Tab~\ref{table:cases}: TM - 2]\\
\end{displayquote}
In extant literature on work motivation, Kahn~\cite{kahn1990psychological} found that the psychological experience that one's work is meaningful contributes to the degree to which people engage in that work and push on in the face of challenges. While work engagement is an individual experience, the team-level equivalent has been called team morale by organizational psychologists~\cite{ivey2015assessment,van2007direct} or \textit{esprit de corps} in military teams~\cite{manning1991morale}. The motivational quality of frequent releases can be understood with goal-setting theory~\cite{locke1981goal} when we consider each release as a tangible and ambitious short-term goal to be achieved by the team. Specifically for Scrum teams, researchers have argued that team members are effectively internal stakeholders and that their satisfaction should be an important goal in addition to customer satisfaction~\cite{mahnic2007using}.

\begin{table}[h]
\small
\centering
\caption{Data structure of observations and themes for the aggregate dimension 'Team Effectiveness'}
\begin{tabularx}{\textwidth}{@{}Xl@{}}
\hline
\textbf{1st Order Observations \& concepts} & \textbf{2nd Order Theme}\\
\hline
Stakeholders are (more) satisfied with the work done by the team& Stakeholder Satisfaction \\
Stakeholders frequently complain about the quality of the product & \\
Stakeholders give compliments to the team for a job well done& \\
Team members feel motivated in their work& Team Morale \\
Team members frequently complain, but there is no conversation about how to solve things & \\
Team members feel like they're in a hamster wheel (different day, same problems) & \\
\hline
\end{tabularx}
\label{tab:datastructureteameffectiveness}
\end{table}

\subsubsection{Responsiveness}
The Scrum teams we observed varied in their ability to respond quickly to emerging needs or issues. This manifested in two related areas. We summarize the observations and second-order themes in Table~\ref{table:datastructureresponsiveness}.

\textbf{Release Frequency: }
The Scrum framework requires Scrum teams to deliver at least one increment every Sprint that can be released to stakeholders~\cite{schwaber2020scrum}. In three cases, teams generally released to stakeholders at least once during the Sprint (Tab~\ref{table:cases}: \#9, \#11, \#12). However, most teams released intermediate versions to internal staging environments (Tab~\ref{table:cases}: \#1, \#2, \#3, \#4, \#5, \#8, \#13) or used several Sprints before they were able to do so (Tab~\ref{table:cases}: \#6, \#10). These teams released batches to their stakeholders relatively infrequently - once per quarter or less - except for critical bug fixes. At the same time, teams recognized the benefits of releasing more frequently. One team stated that:\\
\begin{displayquote}
``It allows us to deliver something to our customers sooner and allows them to suggest changes, or even stop'' [Tab~\ref{table:cases}: TM - 3]\\
\end{displayquote}

While other teams observed that frequent releases contribute to:\\ 
\begin{displayquote}
``rhythm, clarity and getting things done'' [Tab~\ref{table:cases}: TM - 6]\\
``increased software quality'' [Tab~\ref{table:cases}: TM - 6]\\
``happier customers'' [Tab~\ref{table:cases}: SM - 3]\\
\end{displayquote}

We observed that the nature of Sprint Reviews changed as release frequency increased. Teams with high release frequencies (Tab~\ref{table:cases}: \#1, \#2, \#3, \#9, \#11, \#12) inspected the increments in staging or even production environments and used the Sprint Review to collect critical feedback from stakeholders. More stakeholders also attended these sessions. In contrast, teams with lower release frequencies often used static presentations or release logs instead of a working increment to share with stakeholders what was done (Tab~\ref{table:cases}: \#5, \#6, \#10).

\begin{figure*}
\centering
\includegraphics[height=3.5in]{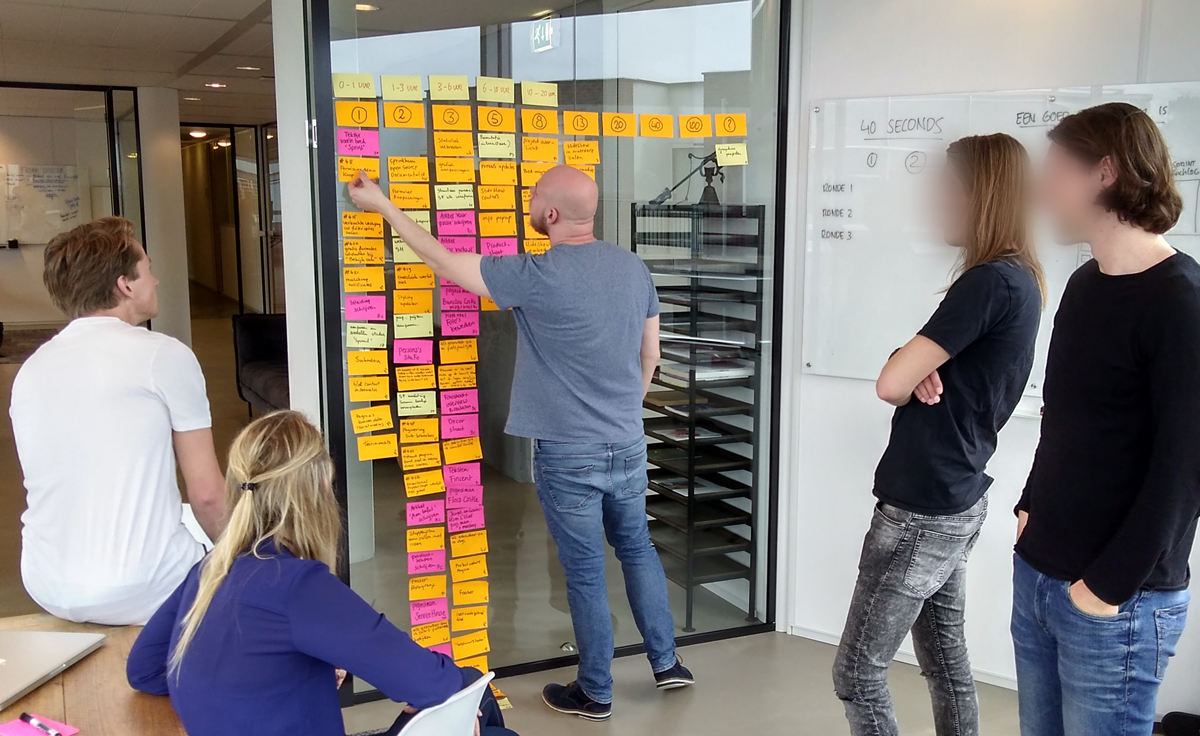}
\caption{A Scrum team engaged in refinement (see Table~\ref{table:datastructureresponsiveness}). Picture by Christiaan Verwijs.}
\label{fig:refinement}
\end{figure*}

\textbf{Refinement:}
The Scrum teams we observed varied in approaching the clarification and decomposition of work on the Product Backlog. This is generally called ``refinement'' in professional literature~\cite{schwaber2020scrum}. Teams engage in this activity to break down large chunks of work into smaller chunks that can be delivered within the scope of a Sprint or less. Most teams performed refinement during weekly workshops attended by the entire team (Tab~\ref{table:cases}: \#1, \#3, \#4, \#5, \#9, see also Figure~\ref{fig:refinement}), which is consistent with other case studies~\cite{paasivaara2016scaling,khmelevsky2017software}. Other teams performed it mainly during Sprint Planning (\#2, \#7, \#8). One consistent pattern was that while most teams performed refinement, they also felt it should be more. For example, one team concluded that:\\
\begin{displayquote}
``We lack the discipline to refine well'' [Tab~\ref{table:cases}: TM - 1]\\
\end{displayquote}

While another Scrum Master observed that:\\

\begin{displayquote}
``When we do refine work, the Sprint seems to be much smoother'' [Tab~\ref{table:cases}: SM - 9]\\
\end{displayquote}

Some teams struggled to see the purpose of refinement. They perceived Sprints as artificial time-boxes where any uncompleted work could be carried over into the next Sprint (Tab~\ref{table:cases}: \#1, \#5, \#7).
In particular, we observed that Sprint Planning tended to go faster for teams that refined during the Sprint. In addition, most Scrum teams frequently raised (the lack of) refinement as a topic during Sprint Retrospectives (Tab~\ref{table:cases}: \#1, \#2, \#3, \#13). This was usually driven by a desire to have more clarity as to what is needed, identify external dependencies earlier, or make Sprint Planning easier. One Scrum Master described the purpose of refinement as:\\

\begin{displayquote}
``a way to identify dependencies on other teams upfront, so that we are not surprised by it during a Sprint'' [Tab~\ref{table:cases}: SM - 1]
\end{displayquote}

\begin{table}[h]
\small
\centering
\caption{Data structure of observations and themes for the aggregate dimension 'Responsiveness'}
\begin{tabularx}{\textwidth}{@{}Xl@{}}
\hline
\textbf{1st Order Observations \& concepts} & \textbf{2nd Order Themes} \\
\hline
The team feels that iterations are artificial (i.e., "There's always a next Sprint" or "Let's just carry it over to the next Sprint") & Refinement \\
Team members spend time during a Sprint to prepare work for the coming Sprints&  \\
Sprint Planning tends to drag on as teams try to discover what they need to create that Sprint &  \\
The Sprint Backlog contains only a handful of large Product Backlog Items &  \\
Sprints generally result in at least one releasable increment for this team& Release Frequency \\
Acceptance testing is done after the Sprint and not during & \\
Sprint Reviews are used to review an Increment that is running on an environment that is as close to production as possible (e.g., not a developers pc)& \\
Sprint Reviews do not demonstrate working software (e.g., instead, PowerPoint presentations are used, or the team simply talks about it) & \\
\hline
\end{tabularx}
\label{table:datastructureresponsiveness}
\end{table}

\subsubsection{Stakeholder Concern}
We found that Scrum teams varied in the degree to which the needs of stakeholders were considered in their decision-making. We define \textit{``stakeholders''} as external parties that have a stake in the outcomes of a team, like customers, key users, and investors. The differences manifested primarily in four areas.

\textbf{Value focus:}
Not all Scrum teams focused equally on the value their work delivered to stakeholders. The Product Owners played an important role here. Some Product Owners frequently discussed with their team how the work on the Product Backlog fit in the overarching strategy (Tab~\ref{table:cases}: \#1, \#3, \#5, \#9, \#12). These teams also periodically removed items from the Product Backlog that no longer fit with that strategy. Three Product Owners also shared how the work returned on investments and how it impacted stakeholders (Tab~\ref{table:cases}: \#1, \#3, \#12). Other Product Owners primarily passed requirements on without elaboration (Tab~\ref{table:cases}: \#6, \#7). We also observed that teams differed in how work was described on the Product Backlog. Some teams (Tab~\ref{table:cases}: \#3, \#6) captured work in technical terms without connecting it to the needs of stakeholders (e.g., ``Update connector API to expose module composition.''). Other teams explicitly clarified how items on their Product Backlog connected to the needs of stakeholders (e.g., ``Administrators can sort submissions by entry-date so they can address the most recent ones first.''). 

The role that Product Owners play in creating a sense of what is valuable is also recognized in extant literature~\cite{bass2016all,unger2021product}. Bass et al. ~\cite{bass2018empirical} concluded from a qualitative study with 46 Product Owners that the role ``is critical for translating business needs into software implementation by gathering and prioritizing requirements.''. Product Owners can use different strategies to promote collective product ownership and develop a shared understanding of value~\cite{judykruminsbeens2008,unger2020product,kristinsdottir2016responsibilities}.

\textbf{Stakeholder collaboration: }
There was a substantial difference between Scrum teams in how they involved stakeholders. Some teams frequently invited stakeholders to their Sprint Reviews (Tab~\ref{table:cases}: \#1, \#2, \#3, \#9). The Product Owner often took the initiative here. Nevertheless, we also observed many teams where developers interacted directly with stakeholders (Tab~\ref{table:cases}: \#1, \#3, \#4, \#7, \#9, \#14). Developers visited stakeholders on-site or contacted them by e-mail or telephone for clarification. In some cases, developers trained stakeholders to use their product (Tab~\ref{table:cases}: \#1, \#2, \#3). In some teams, only the Product Owner interacted with stakeholders (Tab~\ref{table:cases}: \#2, \#5, \#6, \#8, \#11, \#12). The purpose of stakeholder collaboration was mostly described in terms of gathering feedback:\\
\begin{displayquote}
``It allows us to involve users in development to get early feedback'' [Tab~\ref{table:cases}: PO - 7]
``Prioritize work together with all internal stakeholders'' [Tab~\ref{table:cases}: PO - 1]\\
\end{displayquote}

Stakeholder collaboration is recognized as a success factor for Scrum teams in extant literature~\cite{pinton2020human,hoda2011impact,hoda2017becoming,lopez2016problems}. Hoda, Stuart \& Marshall~\cite{hoda2011impact} found that limited collaboration between Agile teams and their stakeholders can lead to unclear requirements, lack of feedback, and loss of productivity and business. Furthermore, less-experienced teams may be more inclined to delegate stakeholder collaboration to their Product Owners~\cite{hoda2017becoming}.

\begin{figure*}
\centering
\includegraphics[height=3.5in]{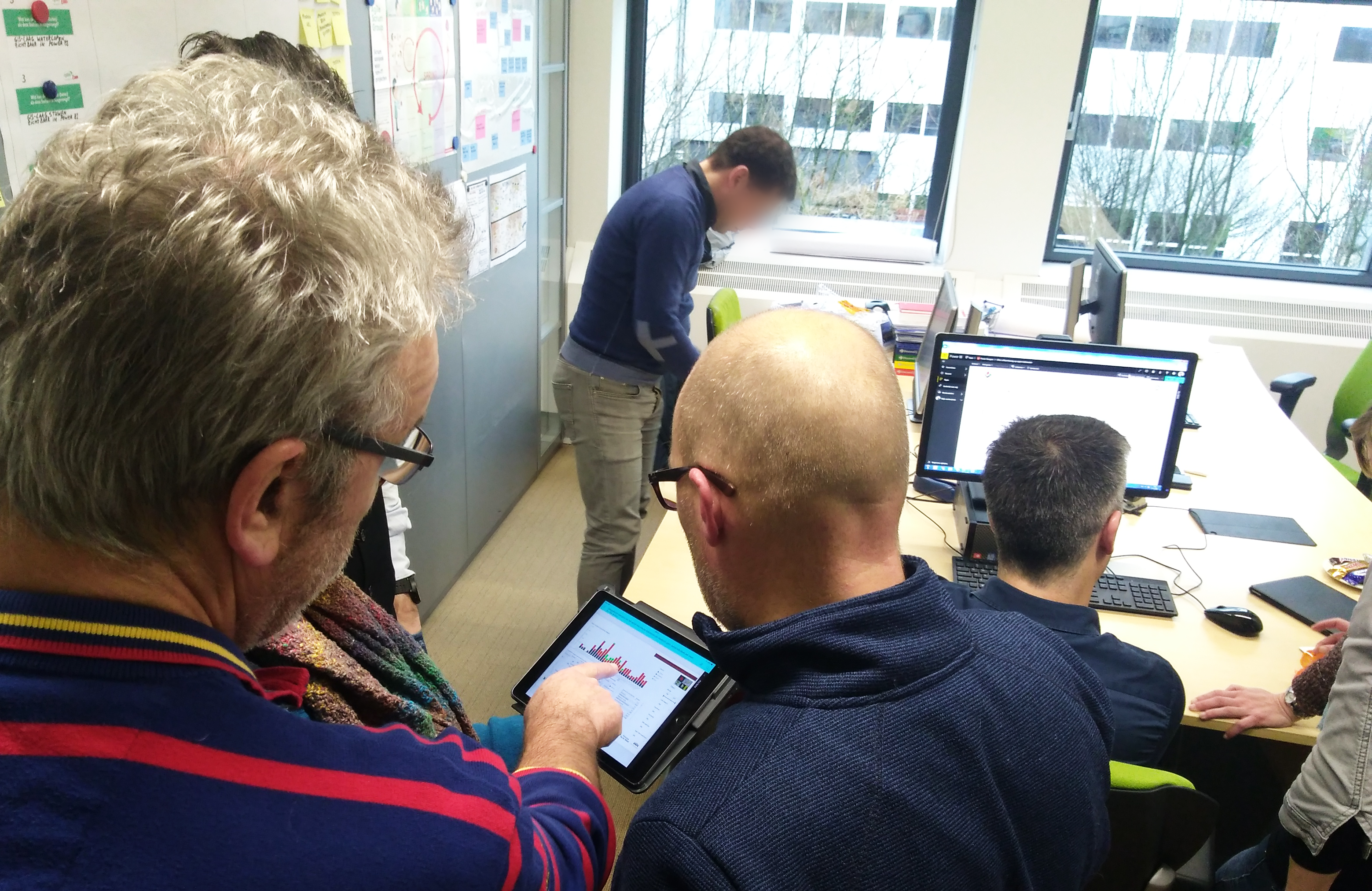}
\caption{Example of an interactive Sprint Review (see Table~\ref{table:datastructurestakeholderconcern}). Picture by Christiaan Verwijs.}
\label{fig:sprintreview}
\end{figure*}

\textbf{Sprint Review Quality: }
The Sprint Review is a formal event in Scrum where feedback is gathered from stakeholders and adjustments are identified~\cite{schwaber2020scrum}. Not all teams used the Sprint Review for this purpose. Some teams never invited stakeholders (Tab~\ref{table:cases}: \#4, \#6, \#7). In the teams that did, the way feedback was gathered differed. Some teams offered a formal presentation (Tab~\ref{table:cases}: \#1, \#6, \#8, \#11). Other teams offered stakeholders opportunities to ``take the wheel'' and try new features on the spot (Tab~\ref{table:cases}: \#3, \#5, \#8, \#9, \#13). This interactive approach tended to draw out more constructive feedback than formal presentations. Only in a few cases did teams use Sprint Reviews to reflect on the larger objective and inform the following steps (Tab~\ref{table:cases}: \#1, \#9). Product Owners were often instrumental in shaping the quality of Sprint Reviews by actively inviting stakeholders (Tab~\ref{table:cases}: \#1, \#9 and \#13) by initiating preparation with the team (Tab~\ref{table:cases}: \#7 and \#9) or by suggesting interactive formats (Tab~\ref{table:cases}: \#9). In one example of an interactive format, the Product Owner set up four stations where stakeholders could try new features with help from team members. Feedback was collected at the stations on postcards as stakeholders rotated across the stations during the Sprint Review (see Figure~\ref{fig:sprintreview}).

We could not find studies that specifically identified Sprint Reviews as a success factor. However, professional literature emphasizes the Sprint Review as a critical opportunity to align the work of a team with its stakeholders ~\cite{verwijs_schartau_overeem_2021,ripley2020fixing,pichler2016strategize}. We speculate that the reason for this gap is that the quality of Sprint Reviews is assumed to result from other factors in this dimension, such as lack of value focus, low collaboration with stakeholders, and unclear goals. However, we observed that some of the teams we studied improved in these areas when they started inviting more stakeholders to their Sprint Reviews.

\textbf{Shared goals: }
Shared goals are used in the Scrum framework to encourage collaboration and clarify how individual tasks connect to larger business goals~\cite{schwaber2020definiton}. In practice, the use of shared goals varied greatly. In some cases, Product Owners began Sprint Planning by stating a business objective that then guided the ordering and selection of work for that Sprint (Tab~\ref{table:cases}: \#1, \#2, \#7, \#8, \#9, \#12, \#13). Other teams selected the top items, even when they were unrelated (Tab~\ref{table:cases}: \#3, \#4, \#5, \#6, \#10). Teams that did not use shared goals often acknowledged that it would bolster team spirit, customer orientation, and collaboration if they would (Tab~\ref{table:cases}: \#1, \#4, \#7).

The importance of shared goals is also recognized in the extant literature on work teams. The presence of salient shared goals positively influences team effectiveness~\cite{aube2005team, daspit2013cross,janz1997knowledge}, cohesion~\cite{man2003effects}, motivation~\cite{kirkman1999beyond} and cooperation~\cite{molm1994dependence}. For Scrum teams specifically, shared goals allow Scrum Teams to align on important business objectives, both within teams and the broader organization~\cite{kristinsdottir2016responsibilities,van2015empirical,whitworth2007social}.

\begin{table}[h]
\small
\centering
\caption{Data structure of observations and themes for the aggregate dimension 'Stakeholder Concern'}
\begin{tabularx}{\textwidth}{@{}Xl@{}}
\hline
\textbf{1st Order Observations \& concepts} & \textbf{2nd Order Theme}\\
\hline
The team frequently spends time cleaning up the Product Backlog by removing items that do not make sense anymore (e.g., in light of changed product vision, strategy) & Value Focus \\
Team members other than the Product Owner can reiterate the strategy behind the upcoming sprints& \\
The team talks about how or why items on the Product Backlog are valuable to their stakeholders during various Scrum Events& \\
The Backlog contains mostly technical tasks and is difficult to connect to specific business needs by stakeholders & \\
The team is familiar with the vision and strategy for the product they are working on& \\
The work that a team does is aligned with other aspects of the business (e.g., support, sales, marketing)& Stakeholder Collaboration \\
Except for the Product Owners, the members of a team do not generally speak with users, customers, or other stakeholders & \\
Product Owners involve customers/users/stakeholders in the prioritization of the Backlog & \\
Product Owners frequently meet with the customers or users using their software& \\
The Backlog is not easily available to stakeholders, end-users, and other teams & \\
There is no interaction during Sprint Reviews between members of the team; people sit back and watch & \\
Stakeholders have opportunities during Sprint Reviews to interact and review a working product or working features& Sprint Review Quality \\
The team prepares for the Sprint Review by determining what they would like feedback on and in what order& \\
The Sprint Review is attended by users, customers, and other stakeholders& \\
The Product Owner uses the Sprint Review to get feedback from stakeholders on the work that was done& \\
Most Sprints have a Sprint Goal that explains how the work in a Sprint tie together& Shared Goals \\
There is a roadmap of Sprint Goals for the upcoming sprints& \\
\hline
\end{tabularx}
\label{table:datastructurestakeholderconcern}
\end{table}

\subsubsection{Team Autonomy}
The observed teams varied in their autonomy in making decisions about how to perform their work. This broadly concerned two types of constraints to their autonomy. The first type was imposed from outside the team (self-management). The second type emerged from skill-based constraints within the team (cross-functionality). We summarize the observations and second-order themes for this aggregate dimension in table~\ref{table:datastructurestakeholderconcern}.

\textbf{Self-Management: }
The ability of the observed Scrum teams to self-manage their work varied by case. For example, the use of the work method was often suggested by management and co-opted by teams after exploratory workshops (Tab~\ref{table:cases}: \#1, \#2, \#5, \#6, \#8, \#9, \#10, \#11, \#12, \#15). During these workshops, the participants distributed roles, scheduled the formal Scrum events, and clarified collective ambitions. Scrum teams controlled their composition in some cases (Tab~\ref{table:cases}: \#1, \#3, \#6, \#14). One department manager initially proposed a composition for five teams and then decided against this during a collective kickoff workshop with everyone present:\\
\begin{displayquote}
``I discovered today that I should not make decisions for others. So I will leave the composition of the teams to you instead'' [Tab~\ref{table:cases}: DM - 13]\\
\end{displayquote}

In another case, a department manager proposed an initial composition for several teams and encouraged them to switch members when they discovered better compositions (Tab~\ref{table:cases}: \#6). While all teams in our study could distribute work internally as they saw fit, their control over incoming work and when to perform it varied greatly. In one case, a Scrum Master observed:\\
\begin{displayquote}
``Our Product Owners have no autonomy. They are told what to do, when to do it, and often how to do it, and they pass that on to the team'' [Tab~\ref{table:cases}: SM - \#10]\\
\end{displayquote}

Similarly, contractual agreements regarding the scope and delivery dates were often controlled by external departments or steering committees (e.g., Tab~\ref{table:cases}: \#1, \#2, \#4, \#7). Only a few teams were directly responsible for their contractual agreements with stakeholders (Tab~\ref{table:cases}: \#3, \#14).

The notion of self-management is central to Scrum. The Scrum framework describes Scrum teams as \textit{``self-managing, meaning they internally decide who does what, when, and how''}~\cite{schwaber2020definiton}. This follows the principle of \textit{responsible autonomy} from sociotechnical systems (STS)~\cite{trist1951some}. In order to make teams more resilient in unpredictable environments where managers cannot constantly regulate work, the team as a whole is made responsible for its composition, task distribution, and leadership. Other studies have also identified high autonomy as an important success factor for Agile teams~\cite{donmez2013practice,f2018Agile,moe2010teamwork}. Self-managing teams have been associated with a range of positive effects by organizational psychologists~\cite{kirkman1999beyond}, such as increased productivity, job satisfaction, and higher commitment.

\textbf{Cross-functionality: }
Another constraint to autonomy emerged from the skills present in teams. Some of the Scrum teams covered a diverse range of skills from testing, business analysis, and design to coding (Tab~\ref{table:cases}: \#1, \#3, \#5, \#6, \#9). The range of skills was much narrower in other teams, often only coding (Tab~\ref{table:cases}: \#2 and \#7). We observed two consequences of low versus high functional diversity. First, when a team lacked a specific skill and was not available in the environment, they would forego tasks altogether (e.g., testing, user interface improvements, security hardening). This impacted the quality and usability of their product (Tab~\ref{table:cases}: \#6 and \#7). Second, when a team lacked the skill, and it was available in their environment, this effectively introduced a dependency on people outside the team who possessed that skill. These dependencies increased development time as teams had to wait for dependencies to complete the task (Tab~\ref{table:cases}: \#1, \#2, \#5, \#6, \#10). For example, teams in case \#1 (Table~\ref{table:cases}) frequently waited for a vendor to deploy new versions to one of a dozen environments. This delayed internal testing and caused confusion as to which version was running on which environments at a given moment in time. Another striking example from this case was how the initial composition of Scrum teams included skills for coding, design, and testing but not requirements analysis. This was initially done by a separate team and created a recurring dependency as Scrum teams ran into questions about the requirements on the one hand and delivered work that did not meet the requirements on the other. This resulted in frequent rework and bugs. When the requirement analysts later joined the Scrum teams, the department manager concluded:

\begin{displayquote}
``this greatly improved quality now that testing is part of the development cycle'' [Tab~\ref{table:cases}: DM - 6]
\end{displayquote}

In extant literature on work teams, Keller~\cite{keller2001cross} describes cross-functional teams as consisting of members with different functional backgrounds, which contributes to \textit{organizational ambidexterity}~\cite{donmez2013practice}. Functional diversity has been linked to shorter development time~\cite{eisenhardt1995accelerating}, increased quality and schedule performance~\cite{keller2001cross} and higher performance~\cite{dechurchmesmermagnus2010,kearney2009and} in general work teams. For Scrum Teams specifically, Moe et al. ~\cite{moe2010teamwork} found that team effectiveness is inhibited when teams lack cross-functionality, as expressed by a high degree of skill specialization and division of work.

\begin{table}[h]
\small
\centering
\caption{Data structure of observations and themes for the aggregate dimension 'Team Autonomy'}
\begin{tabularx}{\textwidth}{@{}Xl@{}}
\hline
\textbf{1st Order Observations \& concepts} & \textbf{2nd Order Theme}\\
\hline
The Scrum Master is distributing work and chairing the Planning and the Sprint Review & Self Management \\
The Product Owner has a mandate to make decisions about the work that the team spends time on&  \\
The team is responsible for planning its work capacity& \\
Teams are involved in or have complete control over decisions that affect their composition or work method & \\
Dependencies slow teams down on external teams or individuals who need to perform tasks that they cannot do themselves & \\
Teams are given responsibility only for specific functional areas (e.g., development, testing, design) & \\
Individual team members are made responsible (and rewarded) for specific specializations & Cross-Functionality \\
Team members collaborate on tasks from the Sprint Backlog, regardless of their specializations& \\
Team members believe they are solely responsible for their specialization (e.g., "I'm only here to code") &  \\
Team members actively keep each other up to date about what they are working on throughout the day& \\
Teams are organized in terms of functional specializations (e.g, a 'continuous delivery team' or 'a design team') &  \\
The testing for a particular item on the Sprint Backlog is performed by someone other than the developer (in or outside the team) & \\
\hline
\end{tabularx}
\label{table:datastructurestakeholderconcern}
\end{table}

\subsubsection{Continuous Improvement}
The teams we studied also varied in the degree to which they engaged in continuous improvement. This manifested primarily in five areas. We summarize the observations and second-order themes for this aggregate dimension in Table~\ref{table:datastructurecontinuousimprovement}.

\begin{figure*}
\centering
\includegraphics[height=3.5in]{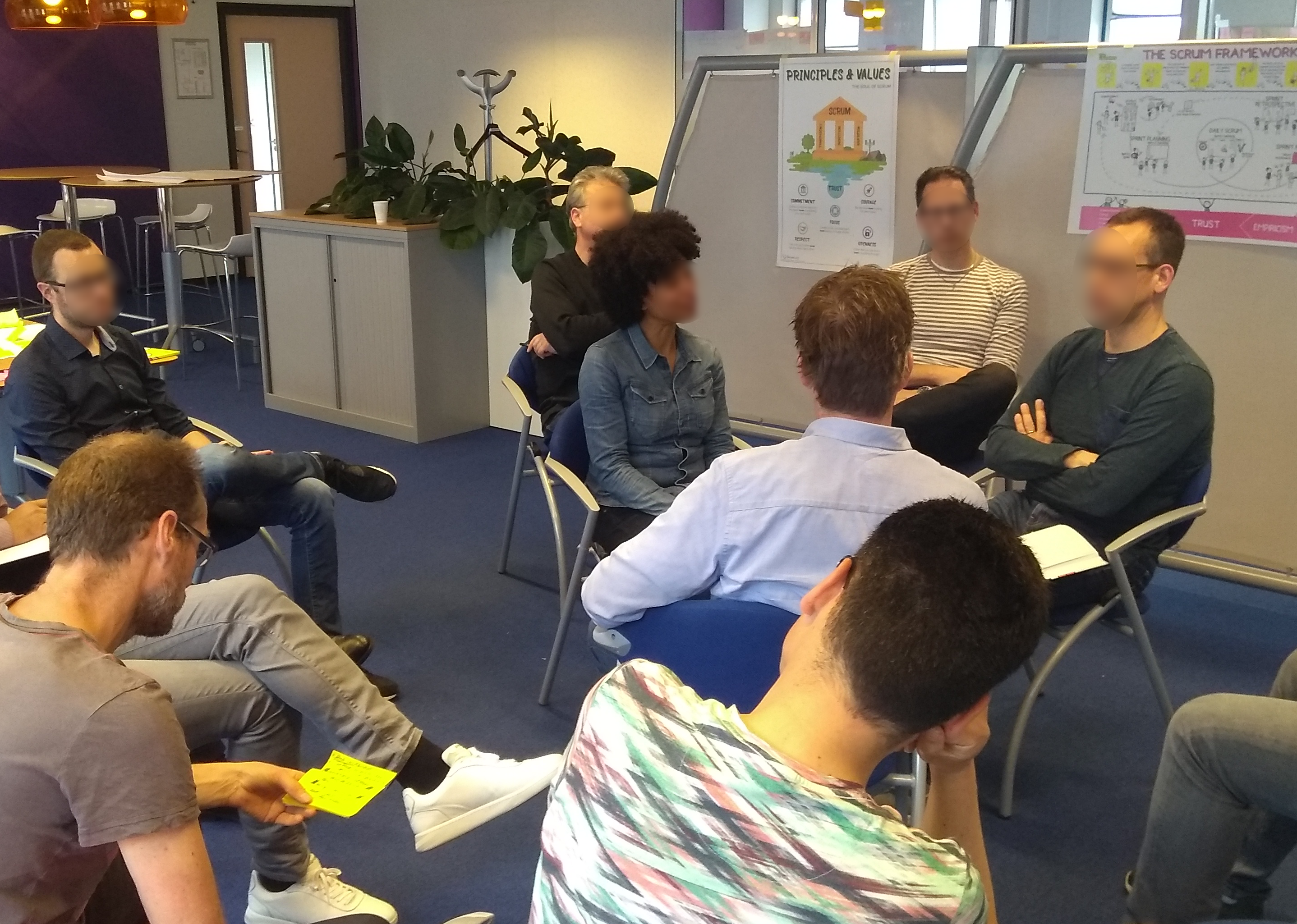}
\caption{Scrum Masters from one organization engaged in a shared learning session (Table~\ref{table:datastructurecontinuousimprovement})}
\label{fig:sharedlearning}
\end{figure*}

\textbf{Sprint-Retrospective Quality: }
The Sprint Retrospective is a recurring opportunity for teams to ``plan ways to increase quality and effectiveness''~\cite{schwaber2020definiton}. The majority of the Scrum teams we observed organized a Sprint Retrospective every Sprint (Tab~\ref{table:cases}: \#1, \#3, \#5, \#6, \#7, \#8, \#9, \#12, \#13). However, other teams did so less frequently (Tab~\ref{table:cases}: \#2, \#4, \#10, \#11). In terms of what was discussed during Sprint Retrospectives, we observed that some teams used different formats to address different themes (e.g., the Definition of Done, customer satisfaction, or code quality)  (Tab~\ref{table:cases}: \#1, \#2, \#3, \#4, \#5, \#6, \#9, \#12). Though a small number of Scrum teams generally repeated the same format, usual variations of a ``plus delta''\footnote{In this format, teams generally create two columns on a shared work-space. One column lists the things that went well (the ``plus''), while another shows improvements (the ``delta'')} to discuss general improvements (Tab~\ref{table:cases}: \#7, \#10). Compared with the repeated use of the same format, we observed that a more diverse range of formats and themes tended to generate more improvements in a broader range of areas (see Figure~\ref{fig:sharedlearning}).

The value of continuous improvement and organizational learning, particularly for complex tasks, is well established in extant literature. With its iterative nature and recurring opportunities for process improvements, the Scrum framework is a continuation of Deming's PDCA-cycle~\cite{deming1986out}. Every Sprint allows teams to detect and correct mismatches in their expectations, tools, processes, quality standards, and collaboration within their team and the broader organization. This is essentially a form of group-level \textit{kaizen}, where groups of employees work together to "find and solve problems faced during their day-to-day work without any interference from management"~\cite{bhuiyan2005overview}. In research on organizational learning, Argyris~\cite{argyris1999organizational} describes this process of detecting errors and correcting them as "learning" and argues that it involves changing actions (single-loop) or challenging existing systems and governing variables (double-loop). Our findings are also congruent with Hoda \& Noble~\cite{hoda2017becoming}, who found that the use of Sprint Retrospectives and other reflective practices became more focused and embedded as teams became more experienced with Agile practices.

\textbf{Concern for Quality}
The Scrum teams we observed also varied in their concern for quality standards. This was most evident in their Definition of Done, which is used in the Scrum framework to clarify the ``quality standards required for the product''~\cite{schwaber2020definiton}. While some teams frequently challenged their Definition of Done to reformulate or add criteria (Tab~\ref{table:cases}: \#1, \#3, \#5, \#8, \#9, \#12, \#12), others never or rarely did (Tab~\ref{table:cases}: \#2, \#6, \#10). One Scrum team did not have an explicit Definition of Done (\#11). The concern for quality was also apparent in how new practices and technologies were explored to improve quality. While some teams frequently did so (Tab~\ref{table:cases}: \#1, \#3, \#5), others rarely did (Tab~\ref{table:cases}: \#2, \#6, \#7, \#10).

We grouped this second-order theme with ``continuous improvement'' because we observed that teams that were highly concerned with quality also tended to engage more in continuous improvement overall. Clear quality standards, as captured in the Definition of Done, provide a clear baseline for improvements and gaps between current and desired quality.

\textbf{Psychological Safety}
We also observed that Scrum teams diverged in how they addressed internal conflicts. For example, in one team (Tab~\ref{table:cases}: \#6), an argument built up between two members over several weeks eventually erupted when one member suddenly left the team, and another ended up at home with a burnout. The Scrum Master observed that:\\
\begin{displayquote}
``[...] people interrupt each other constantly and do not listen'' [Tab~\ref{table:cases}: SM - 6]\\
\end{displayquote}

We observed a similar pattern of interrupting and talking over each other in several other Scrum teams (Tab~\ref{table:cases}: \#1, \#4, \#6, \#7). However, other teams purposefully made time for their members to express personal concerns and frustrations (Tab~\ref{table:cases}: \#3, \#5, \#8, \#9). In one instance, the teams actively sought help from an outside coach to address growing frustrations (Tab~\ref{table:cases}: \#1). In another instance, two team members (\#3) expressed deep concern about whether or not their customers were satisfied with their ability to meet deadlines. The team scheduled a 1-hour session to talk things through. One member of this team explained the benefits of this as:\\
\begin{displayquote}
``It is important for me to talk about what is buzzing around in my head so that I can move on'' [Tab~\ref{table:cases}: TM - 3]\\
\end{displayquote}

These observations are indicative of psychological safety. The organizational psychologist Schein~\cite{schein1992can} conceptualized this as a climate where people can focus on shared goals over self-protection. More recently, Edmondson~\cite{edmondson1999psychological} defined it as \textit{``the shared belief that it is safe to take interpersonal risk''}. Psychological safety has been demonstrated to influence learning behavior, decision quality, and performance in work teams~\cite{edmondson2014psychological}, also specifically in Agile teams~\cite{dreesen2021second,duhigg2016google}. Moe, Dings{\o}yr \& Dyb{\aa}~\cite{moe2010teamwork} conclude that without sufficient trust at the group level, \textit{``team members will expend time and energy protecting, checking, and inspecting each other as opposed to collaborating to provide value-added ideas''}. The ability to give and receive peer feedback is one of the core components of effective teamwork in Agile teams \cite{strode2022teamwork}.

\textbf{Shared Learning}
While Scrum teams can often make substantial improvements to their processes, many improvements are bound to transcend individual teams. The degree to which Scrum teams actively involved others varied greatly. Some teams rarely involved others (Tab~\ref{table:cases}: \#2, \#4, \#7, \#10). However, other teams shared their learning with other teams in a ``marketplace,'' or ``Scrum of Scrums'' (\#1, \#6), invited management (Tab~\ref{table:cases}: \#3, \#8, \#9), organized cross-team workshops on shared challenges (Tab~\ref{table:cases}: \#6, \#12, \#13) or started Communities of Practice (Tab~\ref{table:cases}: \#1, \#10, \#12, \#13).

The need to extend learning beyond team boundaries is also well-established in the extant literature on organizational learning. This process is most commonly referred to as ``boundary crossing''~\cite{kasl1997teams,decuyper2010grasping}. This literature treats teams as open systems that are embedded in larger systems where their effectiveness is negotiated on the team boundaries~\cite{decuyper2010grasping}. Decuyper et al. ~\cite{decuyper2010grasping} argue that ``Teams can neither learn nor work effectively if they hold to share knowledge, competency, opinions or creative ideas across their boundaries with the different stakeholders in the learning process''. Frequent-boundary crossing increases efficiency and innovation~\cite{sundstrom2000work} and increases knowledge transfer~\cite{argote1993group}. The need for inter-team learning is also recognized as an important coordination mechanism in large-scale Agile by Berntzen et al. ~\cite{berntzen2022taxonomy}.

\textbf{Learning Environment}
Finally, we observed that the degree to which the environment of Scrum teams fostered learning directly significantly varied. In some cases, team members were actively encouraged by management to learn new technologies during work hours (Tab~\ref{table:cases}: \#3, \#4, \#5, \#12). In other cases, learning was considered more of a personal responsibility to perform outside of work hours (Tab~\ref{table:cases}: \#8, \#6, \#10). For example, some organizations did not allow team members to participate in professional meetups during work hours (Tab~\ref{table:cases}: \#6, \#10), whereas others actively hosted public, professional meetups during work hours (Tab~\ref{table:cases}: \#2, \#5).

The relevance of a learning environment to continuous improvement is also central in the extant literature on organizational learning. The presence of a culture that nurtures and encourages learning and treats it as an integral part of work is essential for the learning organizations~\cite{argyris1999organizational,marsick1999facilitating}. If learning is not adequately incentivized, it will be hard to develop productive learning behaviors in organizations~\cite{garvin2008yours}.

\begin{table}[h]
\small
\centering
\caption{Data structure of observations and themes for the aggregate dimension 'Continuous Improvement'}
\begin{tabularx}{\textwidth}{@{}Xl@{}}
\hline
\textbf{1st Order Observations \& concepts} & \textbf{2nd Order Themes} \\
\hline
Sprint Retrospectives always yield the same issues over extended periods of time & Sprint Retrospective Quality \\
Sprint Retrospectives are used to dive into deeper problems, not just the superficial symptoms& \\
The action items from Sprint Retrospectives are usually not implemented & \\
A lot of bugs and issues are discovered during the Sprint review or quickly after & Concern for Quality \\
Team members often talk about how to improve the quality of their work (e.g., during Sprint Retrospectives)& \\
The Definition of Done does not define what high-quality looks like for a team & \\
Team members are quick to shoot down ideas (e.g., ``We've tried that before...", ``That doesn't work here") & Psychological Safety \\
Team members generally offer vague and unclear contributions to the Daily Scrum (e.g., ``I've was busy yesterday, I will be busy today") & \\
Team members make an effort to listen to each other before sharing their views & \\
Team members openly ask for help when they are stuck& \\
When a mistake is made, people are not singled out or punished for it& \\
Teams share their learnings and challenges with other teams in the organization& Shared Learning \\
Teams involve people from outside the team to work together to overcome challenges they face& \\
Team members express a desire to try new things or experiment with new technologies& Learning Environment\\
Team members are motivated or encouraged to learn new things, read professional books, visit conferences or meetups& \\
\hline
\end{tabularx}
\label{table:datastructurecontinuousimprovement}
\end{table}

\subsubsection{Management Support}
\label{casestudies:managementsupport}

The role of managers varied greatly between cases. In some cases, managers primarily took a supporting role by asking the teams what they needed from them (Tab~\ref{table:cases}: \#1, \#2, \#3, \#4, \#5, \#7, \#10, \#13, \#14). Management remained distant from the teams in other cases (Tab~\ref{table:cases}: \#9, \#10, \#11) or retained a directive style (Tab~\ref{table:cases}: \#6, \#7). In one extreme example, a top manager would occasionally leave a memo in the team room with sharp criticism about development speed (Tab~\ref{table:cases}: \#6). A Scrum Master in this organization initially described this as:\\
\begin{displayquote}
``[...] management only performs lip service to Scrum, because they are rarely present to help us'' [Tab~\ref{table:cases}: SM - 6]\\
\end{displayquote}

In several cases, we observed how managers supported teams in expanding their skills. This involved training and personal development in some cases (Tab~\ref{table:cases}:  \#3 and \#5), adding members with missing skills (Tab~\ref{table:cases}: \#1, \#2, \#3, \#4), or giving teams the freedom to change their composition as needed (Tab~\ref{table:cases}: \#5, \#6, \#14). We also observed instances where managers actively managed the boundaries of the Scrum teams. For example, in one case (Tab~\ref{table:cases}: \#3), the company owner openly renegotiated the distribution of responsibilities for particular activities (sales, salary setting, company vision) with the team. In another (Tab~\ref{table:cases}: \#1), the program manager moved deployment engineers from an external vendor into each of the teams to remove a dependency that made it difficult for teams to deploy quickly.

Because Scrum teams are self-managing, managers have to shift their approach to what Manz et al. ~\cite{manz1987leading} describe as ``leading others to lead''. The locus of control for work-related decisions shifts from external managers to the teams themselves~\cite{pearce1987design,lee2017self,van2013Agile,de2014Agile}. Russo~\cite{russo2021ASM} investigated success factors for Agile transformations and found that commitment from top management is an important predictor. This requires management to understand why teams work with Scrum, to invest in skills and technologies, and observe the autonomy of teams~\cite{russo2021ASM}. These findings are congruent with research about Agile teams specifically~\cite{van2013Agile,de2014Agile}. Young \& Jordan~\cite{young2008top} provide evidence based on case studies that support from top management may be the most critical factor to project success.

\begin{table}[h]
\small
\centering
\caption{Data structure of observations and themes for the aggregate dimension 'Management Support'}
\begin{tabularx}{\textwidth}{@{}Xl@{}}
\hline
\textbf{1st Order Observations \& concepts} & \textbf{2nd Order Themes} \\
\hline
Management predominantly tells team members how to perform their work, rather than asking where they can support them& Management Support \\
Management understands why this team works with Scrum& \\
Management is available to help teams remove impediments& \\
\hline
\end{tabularx}
\label{table:1}
\end{table}

\section{Theory Development And Hypotheses}
\label{sec:theorydevelopment}
\begin{figure*}
\centering
\includegraphics[height=3.5in]{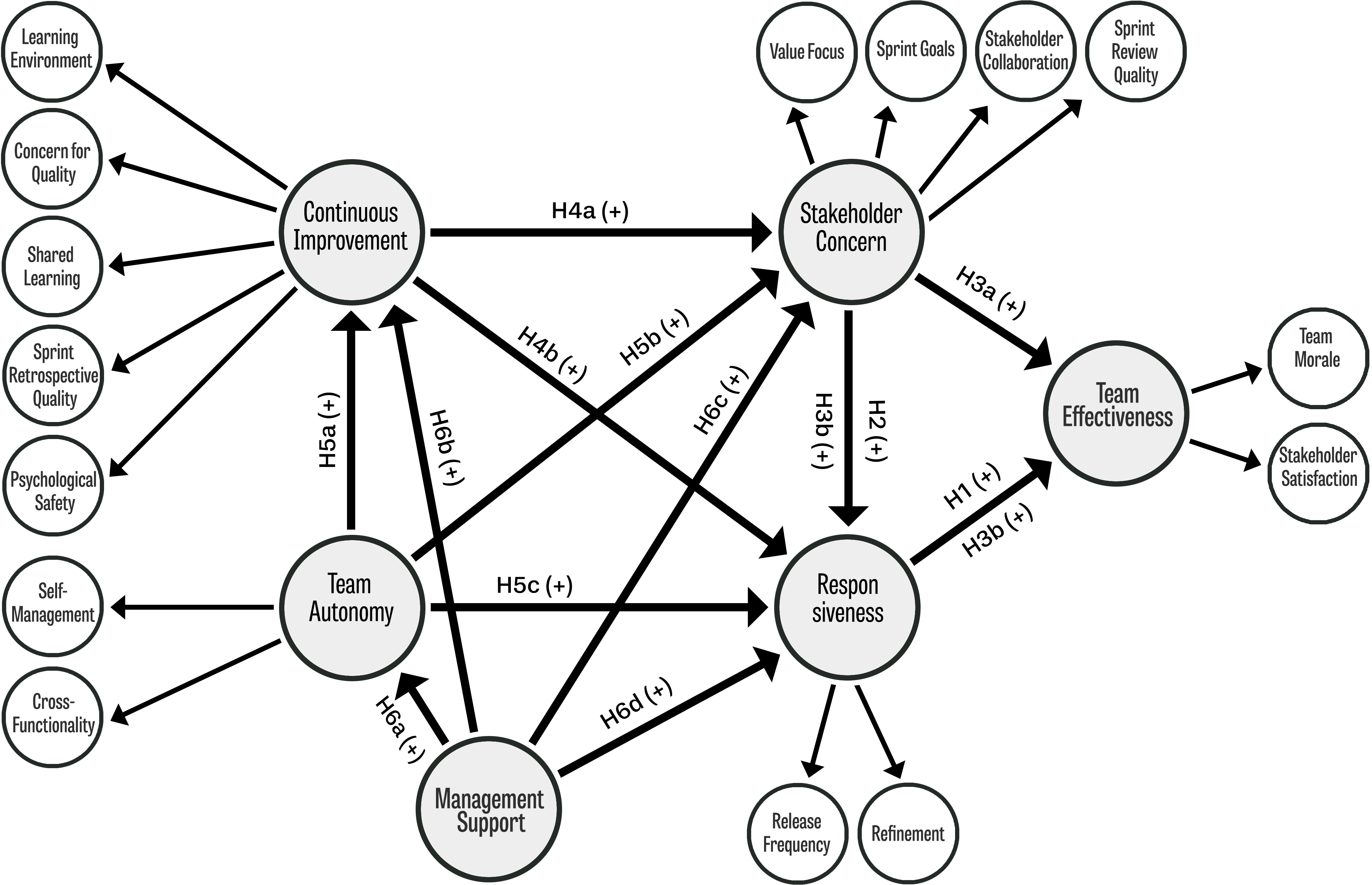}
\caption{Theoretical model and hypotheses.}
\label{fig:hypotheses}
\end{figure*}

The qualitative phase of our study aimed to induce a theoretical model. To answer RQ$_1$, this model should explain how the induced team-level factors influence team effectiveness. Regarding RQ$_2$, we aim to develop a model that can be generalized across settings. However, the exploratory nature of our case studies and their natural research setting~\cite{stol2018abc} are not suited to generalization. This is why the second phase of our study aims to test our theoretical model with quantitative methods and a substantial sample of Scrum teams. We will now present this model and develop testable hypotheses. Our six hypotheses, and their sub-hypotheses, are visualized in figure~\ref{fig:hypotheses}. 

As described in section~\ref{sec:multiplecasestudy}, we induced ``Stakeholder Satisfaction'' and ``Team Morale'' as indicators of team effectiveness. Other outcomes of effective Scrum have been identified in the literature, like increased quality~\cite{cheng2009controlling} and project success~\cite{russo2021ASM}. However, organizational psychologists often define team effectiveness as \textit{``the degree to which a team meets the expectations of the quality of the outcome''}~\cite{hackman1976design}. In this sense, stakeholder satisfaction is the evaluation of team outcomes from the external perspective of stakeholders, whereas team morale is the evaluation of team outcomes from the internal perspective of team members. This is conceptually similar to how team effectiveness is defined in the ``Team Diagnostic Survey (TDS)'' by Wageman, Hackman \& Lehman~\cite{wageman2005team}, and adds increased interdependence between members as a third criterion.
In software development, team effectiveness is often understood more broadly to include developer productivity (e.g., lines of code, hand-offs, merge times, velocity), such as the SPACE framework by Forsgren et al. ~\cite{forsgren2021space}. However, productivity and team performance are highly contextualized. What is considered highly productive in one organization may be low in another. Mathieu et al. ~\cite{mathieu2008team} note that such objective measures are difficult to compare and aggregate across organizations, as we aim to do in this study. So similar to other studies that have investigated team effectiveness~\cite{doolen2003impact,purvanova2021impact,kline1997predicting}, we use the satisfaction of team members and stakeholders as a composite measure of team effectiveness.

We propose that the effectiveness of Scrum teams depends on five team-level factors: (a) the ability of Scrum teams to respond quickly to emerging needs, (b) the awareness that teams have of the needs of their stakeholders, (c) the autonomy that teams have to perform their work as they see fit, (d) a climate of continuous improvement and (e) support from management. Based on the findings from the case studies and supported by extant literature, we propose a causal process by which these factors interact to influence team effectiveness.

One factor that is central to our model is responsiveness. This is the capability of Scrum teams to respond rapidly to changing needs and emerging insights through the frequent release of increments. In the first phase of our research, we identified two relevant indicators for responsiveness: the release frequency of a team and their ability to refine large chunks of work into smaller chunks of work (see section~\ref{sec:multiplecasestudy}). Higher responsiveness is effectively what distinguishes Agile methodologies from more plan-based approaches~\cite{AgileManifesto2001,schwaber1995}. We expect that responsiveness positively influences our indicators of team effectiveness. Regarding stakeholder satisfaction, we expect that more responsive Scrum teams have more satisfied stakeholders. Responsive teams can more quickly respond to emerging needs and deliver valuable outcomes to their stakeholders. Regarding team morale, we expect that more responsive Scrum teams will experience higher team morale. As we also observed in the case studies, frequent releases to stakeholders make the work more fulfilling, purposeful, and motivating for team members.\\

Hypothesis 1 (H1). \textit{The responsiveness of a team is positively associated with team effectiveness.}\\

While responsiveness is a predictor of team effectiveness in our model, we expect responsiveness alone is not enough. It is also essential for Scrum teams to ensure that each release addresses required stakeholder needs. Otherwise, a high release frequency only results in a stream of increments of questionable value to stakeholders. We expect that Scrum teams are most effective when they combine responsiveness with a shared understanding of what their stakeholders need, and not just one or the other. In the qualitative phase of our research, we defined this as ``Stakeholder Concern'' and identified Sprint Reviews, stakeholder collaboration, value focus, and shared goals as relevant indicators. ``Stakeholder Concern'' can be thought of as a collection of team mental models about what is valuable to stakeholders from the perspective of team cognition~\cite{cannonbowerssalas1993, roussemorris1986}. Cannon-Bowers \& Salas~\cite{cannonbowersjanissalas2001} argue that team mental models contribute to behavioral coordination on interdependent tasks, the motivational state and cohesion of its members, and their ability to meet objectives. Team cognition has been shown to explain a significant amount of variance in team performance~\cite{dechurchmesmermagnus2010}. Specifically for Scrum teams, Moe, Dings{\o}yr \& Dyb{\aa}~\cite{moe2010teamwork} report that a lack of shared mental models about desirable team outcomes impedes cross-functionality and communication. We observed that Product Owners play a critical role in developing this shared understanding, which is supported by literature~\cite{bass2018empirical,bass2016all,unger2021product}. 
Product Owners can use different strategies to promote collective product ownership ~\cite{judykruminsbeens2008,unger2020product,kristinsdottir2016responsibilities}
The use of shared goals to clarify business objectives is one way~\cite{kristinsdottir2016responsibilities}. There is extensive evidence for the positive influence of clear, shared goals on team effectiveness~\cite{aube2005team, daspit2013cross,janz1997knowledge}, cohesion~\cite{man2003effects}, motivation~\cite{kirkman1999beyond} and cooperation~\cite{molm1994dependence}. 

We expect that stakeholder concern has two complementary effects on team effectiveness. The first is that Scrum teams that have a deep concern for stakeholder needs are more likely to see the benefit of releasing more frequently to those stakeholders and are thus more inclined to be responsive:\\

Hypothesis 2 (H2). \textit{Stakeholder concern is positively associated with responsiveness.}\\

Second, we hypothesize that stakeholder concern is positively associated with our two indicators of team effectiveness. We observed in the qualitative phase of our research that teams found it easier to satisfy the needs of stakeholders when they understood those needs clearly. This also mirrors findings by Hoda, Stuart \& Marshall~\cite{hoda2011impact}. Regarding team morale, we expect that a genuine concern for stakeholder needs bolsters the sense of purpose of a team. This finding is well-grounded in positive psychology~\cite{maslow1981motivation,alderfer1972human}. The psychological experience that one's work is meaningful encourages people to engage in that work and push through even in the face of challenges~\cite{kahn1990psychological,schaufeli2002measurement,ivey2015assessment,van2007direct}. 

In addition to this direct effect on Scrum team effectiveness, we also expect to see a mediating effect on responsiveness. This reflects a core principle of Agile methodologies that a strong focus on stakeholder needs has to be coupled with the capability to respond to those needs. Otherwise, it essentially remains a waterfall approach where stakeholders have to trust in a good outcome at the end of the development cycle~\cite{AgileManifesto2001}. We expect that Scrum teams are most effective when they release frequently \textit{and} focus on the needs of stakeholders.\\

Hypothesis 3 (H3). \textit{Stakeholder concern is positively associated with team effectiveness (H3a), and this relationship is further mediated by responsiveness (H3b)}\\

Our first three hypotheses capture how responsiveness and stakeholder concern interact at the team level to influence team effectiveness. Informed by the case studies and extant literature, we identified three additional team-level antecedents: a climate of continuous improvement, team autonomy, and management support.

We begin with the first antecedent, continuous improvement. Every Sprint offers opportunities to detect mismatches in tools, processes, and quality standards and adjust accordingly. This process of detecting errors and correcting them is what Argyris~\cite{argyris1999organizational} defines as ``learning''. We observed that Scrum teams varied in the degree to which they engaged in continuous improvement, as indicated by the quality of Sprint Retrospectives, psychological safety, quality standards to guide improvement, and shared learning with others in the organization. These factors are reminiscent of work by Marsick \& Watkins~\cite{watkins1996action,marsick1999facilitating} on learning organizations, which they characterize as having ``total employee involvement in the process of collaboratively conducted, collectively accountable change directed towards shared values or principles”. Out of the seven dimensions they identify in the ``Dimensions for Learning Organizations Questionnaire'' (DLQP), they include a climate of openness and safety, recurring opportunities for learning, shared learning, and a collective vision to guide improvements. Hoda \& Noble ~\cite{hoda2017becoming} also emphasizes the role of effective learning as a means for Scrum teams to overcome obstacles. We expect that Scrum teams that engage in continuous improvement are more likely to have overcome barriers to responsiveness and stakeholder concern and are thus more effective as a result.\\

Hypothesis 4 (H4). \textit{Continuous improvement is positively associated with stakeholder concern (H4a) and responsiveness (H4b).}\\

The second antecedent concerns team autonomy. In the case studies, we observed that team autonomy was expressed in autonomy from external constraints (self-management) and autonomy due to fewer internal skill constraints (cross-functionality). Consistent with self-determination theory by Deci \& Ryan~\cite{deci2000and}, we expect that the shift from an external to an internal locus of control encourages more proactive behavior in teams towards achieving shared goals while also making them feel more responsible for the outcomes. Specifically, we hypothesize that high-autonomy teams are more inclined to engage in continuous improvement than low-autonomy teams. Many studies have shown that job autonomy increases proactive behavior to improve the work environment~\cite{anand2012job,parker2010making,tripp2016job}. Specifically for Agile teams, Junker et al. ~\cite{junker2021Agile} found that teams are more likely to initiate changes and improvements when autonomy is high. Similarly, we hypothesize that the increased sense of responsibility for their outcomes will lead high-autonomy teams to show greater concern for the needs of stakeholders and more proactive behaviors aimed at understanding those needs. Finally, we expect that high-autonomy teams are more responsive than low-autonomy teams due to decreased external dependencies.\\

Hypothesis 5 (H5). \textit{Team autonomy is positively associated with continuous improvement (H5a), stakeholder concern (H5b), and responsiveness (H5c).}\\

Finally, we learned from the case studies how support from management varied between Scrum teams. Observations from the case studies also suggest that management support is a critical success factor. This is consistent with other studies that have investigated how management supports contributes to the success of Agile adoption ~\cite{russo2021ASM,van2013Agile,boehm2005management} and change initiatives in general~\cite{young2008top}. Hoda \& Noble~\cite{hoda2017becoming} conceptualize the required change in management approach during Agile transitions as shifting from ``driving'' to ``empowering''. Rather than making decisions for teams, management ensures that teams have what they need to make decisions for themselves. Specifically for Scrum teams, we expect that support from management manifests through the other success factors in our model. So management contributes to team effectiveness by removing barriers to autonomy, continuous improvement, stakeholder concern, and responsiveness.\\

Hypothesis 6 (H6). \textit{Management Support is positively associated with team autonomy (H6a), continuous improvement (H6b), stakeholder concern (H6c), and responsiveness (H6d).}\\

\section{Phase II: Large-scale cross-sectional study}
\label{sec:SEM}

After the induction of the theoretical model described in Section~\ref{sec:multiplecasestudy}, we will now evaluate it to assess its generalizability. 

\subsection{Collection of Quantitative Data}
We performed our data collection process through a customized online survey between August 2020 and October 2021 \footnote{The survey is available at the following URL: www.scrumteamsurvey.org.}. The survey has been designed so that Scrum teams can self-assess their development process. The team member visiting the survey website would first complete the survey for their team and then invite other members to participate anonymously through a unique link for their team. This allowed us to aggregate the results from individual participants to the team level. Upon completion of the survey, we provided teams with an automatically generated and anonymized report of their results, allowing them to take concrete actions for improvements.
We operationalized the sixteen second-order themes identified in the multiple case studies with Likert scales, ranging from 1 (Completely disagree) to 7 (Completely agree). We also added scales to control for common method bias and to control for the impact of Covid-19. The instrument is available in Table~\ref{tab:survey} in the Appendix.
We adapted established scales to Scrum teams from existing literature where possible. Alternatively, we created new scales from first-order concepts we collected in the multiple case study. For example, we operationalized the independent variable \textit{Sprint Review Quality} with two questions: ``The Product Owner of this team uses the Sprint Review to collect feedback from stakeholders'' and ``During Sprint Reviews, stakeholders frequently try out what this team has been working on during the Sprint''. As another example, the dependent variable \textit{Stakeholder Satisfaction} was measured with the questions ``Stakeholders frequently compliment this team with their work.'', ``Stakeholders are generally happy with the software this team delivers.'', ``Stakeholders are generally happy with how fast this team responds to their needs.'' and ``Our stakeholders compliment us with the value that we deliver to them.''. An overview of the scales is shown in Table~\ref{tab:scales}. The complete questionnaire is available in Table~\ref{tab:survey} in the Appendix. To improve the usability and measurement reliability of the sample study, we performed three survey pilots between June 2019 and June 2020. 

After the pilots, we employed three strategies to reduce response bias. 
First, we incentivized participation by offering participants a personalized report to support their teams. The report included team scores and a benchmark derived from the overall and aggregated team responses.
In addition, teams received actionable improvement suggestions based on their scores drawn from a professional book~\cite{verwijs_schartau_overeem_2021}.
Second, we sent reminders to participants who started the survey but did not complete it within two days. 
The survey was advertised on channels that are frequented by people interested in Scrum, ranging from industry platforms, blog posts, podcasts, and videos by industry influencers\footnote{Due to the opt-in nature of the survey, we were not able to compute a meaningful response rate. However, we were able to establish that 42\% of the participants who started the survey completed it.}.
Third, we strongly emphasized the anonymous nature of the survey. Although participants could leave their e-mail addresses to receive research updates, this was not required. The personalized report we provided to teams only showed team-level averages when a sufficient number of members participated but never individual results.

In the pilot studies, we observed that many participants first tested the survey with a quick test run with fake answers and then returned for an actual attempt. To prevent such responses from influencing our results, we added an item at the start of the survey to ask respondents if their answers reflected a real team or were meant to test the survey.

The use of a single method - like a survey - introduces the potential for a systematic response bias where the method itself influences answers~\cite{podsakoff2003common}. To control for such common method bias, the most objective approach in current literature is the use of a marker variable that is theoretically unrelated to other factors in the model~\cite{simmering2015marker}. 
We included three items from the social responsibility scale (SDRS5)~\cite{hays1989five} and found a small but significant unevenly distributed response bias. Following recommendations in the literature, we retained the marker variable ``social desirability'' in our causal model to control for common method bias~\cite{simmering2015marker}.

\subsection{Analysis of Quantitative Data}
For our sample, we began with all individual responses from June 3, 2020, and onward ($N=5,324$). 
Because our survey was public, we applied several strategies to remove careless responses from the sample and prevent them from biasing the results~\cite{meade2012identifying}. 
We first discarded 145 responses where the team's name indicated a test, e.g., ``Fake'' or ``Test.'' 
To further reduce the impact of careless responses, we removed 239 participants that completed the survey below the 5\% percentile (6.87 minutes) or entered only a few questions.
We identified a handful of multivariate outliers based on their Mahalanobis distance but did not remove them. These cases may represent unusual cases, and their removal should be done with caution~\cite{de2014applications}.
Our final sample contained 4,940 individuals from 1,978 teams. The composition of the sample is detailed in Table~\ref{tab:samplecomposition}.
Because we wanted to analyze our measures at the team level, we calculated a team level mean for each item in the survey by aggregating it to a unique key that was assigned to each team. Such aggregation is justified when at least 10\% of the variance exists at the team-level~\cite{hair2019multivariate}. We tested this by calculating the team level's Intraclass Correlation (ICC). This measure describes the resemblance of individuals within a higher-level grouping~\cite{hair2019multivariate}. We found that 51\% of the variance existed at the team level, which exceeded the required threshold of 10\% suggested by Hair et al.
Accordingly, we ran a post-hoc power analysis using G*Power~~\cite{faul2009statistical} version 3.1. Such a test allows researchers to determine the probability of correctly rejecting the null hypotheses, given the sample size and the effect size ($f$). We determined that the sample size allows us to correctly capture small effect sizes ($f=.02$) with a statistical power of almost 100\% ($1-\beta= .999$).
In other words, we are very confident that our sample is big enough to provide a reliable outcome of our SEM analysis.

Next, we tested our data for necessary statistical assumptions to run our SEM analysis. We assessed normality by inspecting the skewness and kurtosis of individual items. The skew for all items remained well below the recommended thresholds for kurtosis ($<3$) and skew ($<2$)~\cite{de2014applications}. We tested linearity by entering all pairs of independent and dependent variables into a curve estimation~\cite{gaskin2012data} to see if the relationship was appropriately linear, which was significant in all cases ($p < 0.01$). We assessed homoscedasticity by inspecting the scatter plots for all pairs of independent and dependent variables for inconsistent patterns but found none. Finally, multicollinearity was assessed by entering all independent variables one by one into a linear regression~\cite{gaskin2012data}. The Variance Inflation Factor (VIF) remained below the critical threshold of 10~\cite{hair2019multivariate} for all measures but indicated some multicollinearity of Psychological Safety with other independent variables (between 7.38 and 9.81).

Finally, we identified the most appropriate strategy for dealing with missing data. Unless data is missing completely at random (MCAR), any patterns in missing data may bias the results of multivariate analysis~\cite{hair2019multivariate}. For this, we calculated Little's MCAR test. This is a Chi-Square test that compares the observed patterns of missing data with the patterns that would be expected from a process that results in random missing data~\cite{hair2019multivariate}. Our test showed that data wasn't completely missing at random  ($Chi^2=13,200.799$, $df=9,853$, $p<0.001$)~\cite{hair2019multivariate}. A closer inspection of patterns in the missing data revealed that the percentage of missing data was below 2\% for most items but slightly higher (up to 3.8\%) for four items that measured aspects relating to stakeholders. Overall, missing data remained below the recommended threshold of 10\%~\cite{hair2019multivariate}. Because the percentage of missing data was modest, and to prevent list-wise deletion of cases and lost information, we performed EM maximum likelihood imputation in SPSS~\cite{byrne2010structural}.

We analyzed the data through Structural Equation Modeling (SEM) with the AMOS software package~\cite{arbuckle2011ibm} to effectively test the multiple relationships between our independent and dependent variables at the same time~\cite{hair2019multivariate,byrne2010structural,russo2021PLS}. 
Our theoretical model is composed of a structural (or inner) component to describe as many relationships between independent and dependent variables as the researcher is looking for and a measurement model that allows the inclusion of multiple indicators to measure latent factors and account for measurement error~\cite{kline2015principles}. 
It is important to note that this makes SEM an inherently confirmatory approach that combines multiple linear regressions and confirmatory factor analysis (CFA) with Maximum Likelihood estimation (ML) to produce more consistent and less biased estimates than those derived through Ordinary Least Squares (OLS) that is typically used in multiple regression and ANOVA~\cite{hair2019multivariate}. 
Researchers evaluate their model through several ``Goodness of Fit'' indices and individual paths' statistical significance and effect sizes. The aim is to arrive at a model as parsimonious as possible while also providing a good fit (and thus explanatory power). We discuss the fit indices used in this paper in section~\ref{sec:modelfitevaluations}.

Based on our theoretical model, we created a full latent variable model that contained both the measurement model and the structural model. The measurement model defines relationships between indicator variables (survey items) and underlying first-order latent factors and effectively acts as a CFA-model~\cite{kline2015principles}. The structural model defines the hypothesized relations between latent variables and is effectively a regression model. This approach makes the results less prone to convergence issues because of low indicator reliability and offers more degrees of freedom to the analysis compared to a non-latent model~\cite{harring2012comparison}. Team effectiveness was modeled as a latent factor and measured through team morale and stakeholder satisfaction. Similarly, we modeled the five team-level factors from the case studies as latent factors, each measured through the associated themes that we identified in the case studies. In the resulting path model, team effectiveness is effectively a dependent variable with the other factors as independent variables that directly or indirectly influence it.

Following the approach outlined in literature~\cite{byrne2010structural,russo2021PLS,hair2019multivariate}, we entered all items from our survey as indicator variables into the model, with each indicator variable loading on a single first-order latent factor that represented the associated (scale) construct and first-order latent factors loading on the five proposed second-order factors. Then, we used AMOS to calculate factor scores and errors for each latent factor based on their indicators.

As recommended by the literature, we first evaluated and fitted the measurement model and then introduced the structural part of the model~\cite{kline2015principles,byrne2010structural}.

\begin{table}[!ht]
\centering
\small
\caption{Scales used in the survey study, along with attribution, number of items and reliability (Cronbach's Alpha) based on respondent-level response data ($N=4,919$)}
\label{tab:scales}
\begin{tabularx}{\textwidth}{@{}lXll@{}}
\toprule
\textbf{Construct variable} & \textbf{Items adapted from} & \textbf{\# Items} & $\bm{\alpha}$
\\ \midrule
Concern for Quality & Created by authors from our case studies & 2 & .786 \\
Shared Learning & Created by authors from our case studies & 3 & .790 \\
Learning Environment & Adapted from 'Continuous Learning' scale in DLOQ~\cite{marsick2003demonstrating} & 2 & .725 \\
Sprint Retrospective Quality & Created by authors from our case studies & 3 & .834 \\
Psychological Safety & Adapted from 'Inquiry \& Dialogue' scale in DLOQ~\cite{marsick2003demonstrating} & 5 & .862 \\
Shared Goals & Adapted from Van der Hoek, Groeneveld \& Kuipers~\cite{van2018goal} & 2 & .924 \\
Stakeholder Collaboration & Created by authors from our case studies & 3 & .852 \\
Sprint Review Quality & Created by authors from our case studies & 2 & .691 \\
Value Focus & Created by authors from our case studies & 3 & .799 \\
Cross-Functionality & Adapted from Edmondson~\cite{edmondson1999psychological} & 2 & .735 \\
Self-Management & Adapted from Langfred~\cite{langfred2005autonomy} & 3 & .803 \\
Refinement & Created by authors from our case studies & 3 & .820 \\
Release Frequency & Created by authors from our case studies & 2 & .860 \\
Management Support & Created by authors from our case studies & 2 & .852 \\
Team Morale & Adapted from Van Boxmeer et. al.~\cite{van2007direct} and Schaufeli~\cite{schaufeli2002measurement} & 3 & .894 \\
Stakeholder Satisfaction & Created by authors from our case studies & 4 & .842 \\
Social Desirability & Highest-loading items from SDRS-5 scale~\cite{hays1989five} & 3 & .688 \\
Covid-19 Impact ($N=928$) & Created by authors & 2 & .922 \\ \bottomrule
\end{tabularx}
\end{table}

\subsection{Model fit evaluations}
\label{sec:modelfitevaluations}
Before testing for the model fit, 
we assessed reliability, convergent, and discriminant validity for the resulting measurement model. 
The individual steps involved in the model-fitting process are included in Table~\ref{tab:modelfitting} in the Appendix. 
Discriminant validity was assessed by analyzing the heterotrait-monotrait ratio of correlations (HTMT) with a third-party plugin in AMOS~\cite{gaskin2016master} and following the approach outlined in literature~\cite{hair2019multivariate,henseler2015new}. This ratio between trait correlations and within trait correlations should remain below $R=.90$ to indicate good discriminant validity from other constructs in different settings. This was the case for all measures. 
We assessed convergent validity by inspecting composite reliability (CR) and average extracted variance (AVE). The AVE remained above the rule of thumb of $>.50\%$~\cite{hair2019multivariate} for all pairs of factors, ranging between .583 and .875. The CR was equal to or above the threshold of .7~\cite {hair2019multivariate} for all scales, except Sprint Review Quality ($CR=.689$). We decided to retain this scale due to the modest violation and its theoretical relevance.

We then proceeded with the fitting procedure. AMOS suggested three covariances between items from the same scales (PS2 and PS3, PS1 and PS4, VF2 and VF3) based on high modification indices. A closer inspection showed that the pairs used wording more similar than other items of the same scales, so we allowed the covariances. Finally, we investigated local fit by inspecting the residual covariance matrix. A standardized residual covariance is considered large when it exceeds 2.58~\cite{byrne2010structural}. This indicates that an item does not sufficiently measure (only) its intended factor. We found two items, SH1 and PS5, with multiple violations and removed them from the model. No further modifications were made. 

We then evaluated the goodness of fit using indices recommended by recent literature ~\cite{kline2015principles,byrne2010structural,hair2019multivariate}; the Comparative Fit Index (CFI), the Root Mean Error of Approximation (RMSEA), the Standardized Root Mean Residual (SRMR) and the Tucker Lewis Index (TLI). A commonly used index that we reported but did not test for was $\chi^2$ (CMIN) and its corollary CMIN/df. These indices are highly susceptible to type I errors in larger samples ($N>400$,~\cite{hair2019multivariate}). So instead, we used the Comparative Fit Index (CFI)~\cite{bentler1990comparative} which offers a similar test but with consideration of the sample size and its reliable properties have made it the most commonly used index today~\cite{hair2019multivariate}. A cut-off value of .95 or higher is generally considered to indicate good fit~\cite{byrne2010structural,hair2019multivariate,bentler1980multivariate}. The Root Mean Error of Approximation (RMSEA) by Steiger \& Lind~\cite{steiger1980statistically} also provides an index that considers sample size but adds to this a parsimony adjustment that leads it to favor the simplest model out of potential models with the same explanatory power~\cite{kline2015principles}. A value below .05 is generally considered to indicate a good fit~\cite{byrne2010structural, hair2019multivariate}; additionally, we follow the advice to report the confidence interval in addition to only the absolute value~\cite{chen2008empirical}. The Standardized Root Mean Residual (SRMR) calculates a standardized mean of all the differences (residuals) between each observed covariance and the hypothesized covariance between variables~\cite{hair2019multivariate}. A value below .05 is indicative of a good fit. We also inspected local fit by looking at the standardized residuals between pairs of variables, with values beyond 2.58 as a cut-off value for poor local fit~\cite{byrne2010structural}. Finally, we report and test the Tucker Lewis Index (TLI). This is another incremental fit index, like the CFI, that compares the relative improvement of the hypothesized model from a model where all variables are uncorrelated. Hair et al. ~\cite{hair2019multivariate} considers a value of .92 or above sufficient to conclude a good model fit. In addition to overall model fit, we also evaluated our model on the percentage of variance that is explained in team effectiveness by all other variables in the model. 

We also report a very good model fit of our data ($Chi^2(763)=2.324,465$; $TLI=.963$; $CFI=.970$; $RMSEA=.032$; $SRMR=.028$). A Confirmatory Factor Analysis (CFA) is reported in the Appendix (Table \ref{tab:cfa}) that shows that all items load primarily on their intended second-order factors. The cumulative Eigenvalues of 17 factors explain 79\% of the total observed variance, which is well beyond the recommended threshold of 60\%~\cite{hair2019multivariate}. 

We then proceeded to test the path model for the effects we predicted from our theory. 
Our hypothesized theoretical model fits the data well on each of the fit indices, as described in Table~\ref{tab:modelfitindices}: $Chi^2(710)=2,757.805$; $TLI=.953$; $CFI=.959$; $RMSEA=.038$; $SRMR=.035$. The predictors in our model explain respectively 57.9\% of the variance in stakeholder satisfaction and 34.9\% of the variance in team morale. For studies in the social sciences, values above 26\% are considered large~\cite{cohen1992power}. 

All steps in the fitting process are summarized in Table~\ref{tab:modelfitting} in the Appendix.

\begin{figure*}
\centering
\includegraphics[height=3.5in]{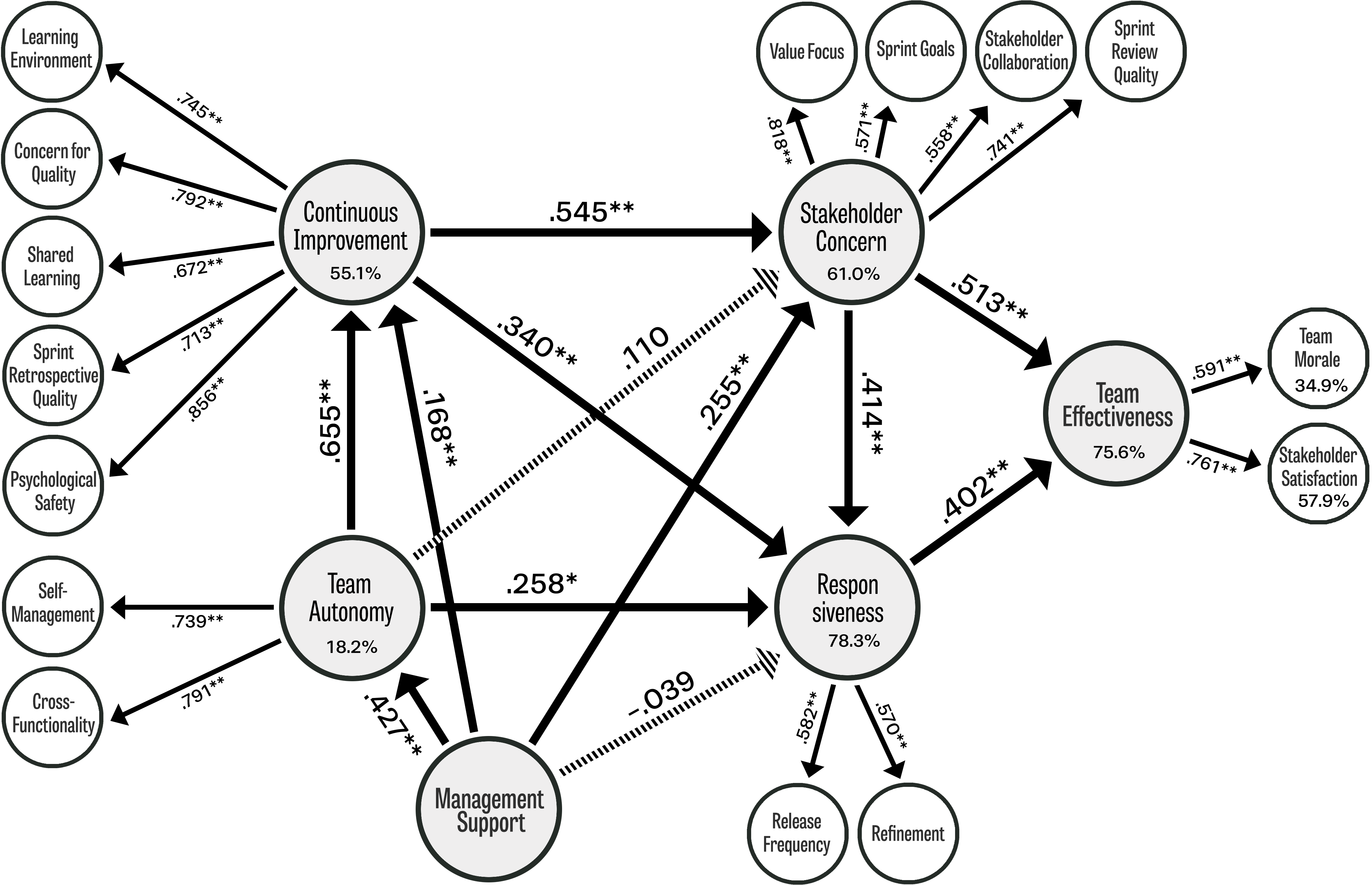}
\caption{Standardized factor loadings and standardized path coefficients for the research model ($**: p < .01$, $*: p < 0.05$). Squared Multiple Correlations are reported at the top right of each endogenous variable. The model shows first- and second-order latent factors. The indicator items are excluded from the model for the sake of clarity. }
\label{fig:SEM}
\end{figure*}

\begin{table}[!ht]
\centering
\small
\caption{Composition of the sample}
\label{tab:samplecomposition}
\begin{tabularx}{0.75\textwidth}{@{}lll@{}}
\toprule
\textbf{Variable} & \textbf{Category} & \textbf{N (\%)} \\ \midrule
Respondents &  & 4,940 \\ 
Teams &  & 1,978 \\ 
Respondent role & Scrum Master & 1,216 (24.6\%) \\ 
 & Product Owner & 465 (9.4\%) \\ 
 & Developer (inc tester, analyst, design) & 2,631 (53.6\%) \\ 
 & Other & 601 (12.2\%) \\ 
 & Unknown & 27 (0.2\%) \\ 
Respondents per team & 1 respondent & 1,366 (69.1\%) \\ 
 & 2-4 respondents & 255 (11,4\%) \\ 
 & 5-8 respondents & 294 (14.9\%) \\ 
 & 9+ respondents & 23 (4.7\%) \\ 
Product Type & Product for internal users & 1,057 (53.4\%) \\ 
 & Product for external users / customers & 894 (45.2\%) \\ 
Scrum Team Size & 1-4 members & 93 (4.7\%) \\ 
 & 5-10 members & 1,404 (71.0\%) \\ 
 & 11-16 members & 364 (4.9\%) \\ 
 & \textgreater{}16 members & 60 (5.1\%) \\ 
 & Unknown & 27 (1.4\%) \\ 
Team Experience & No experience with Scrum & 79 (4\%) \\ 
 & Low experience with Scrum & 458 (23.2\%) \\
 & Moderate experience with Scrum & 1,000 (50.6\%) \\ 
 & High experience with Scrum & 345 (17.4\%) \\ 
 & Unknown & 91 (4.6\%) \\ 
Organization Sector & Technology And Telecommunications & 745 (37.7\%) \\ 
 & Financial & 380 (19.2\%) \\ 
 & Government & 106 (5.4\%) \\ 
 & Other & 747 (37.8\%) \\ 
Organization Size & 1-50 employees & 184 (9.3\%) \\ 
 & 51-500 employees & 537 (27.1\%) \\ 
 & 501-5,000 employees & 679 (34.3\%) \\ 
 & \textgreater{}5,000 employees & 499 (25.2\%) \\ 
 Region & Europe & 1,309 (66.2\%) \\ 
 & North America & 253 (12.8\%) \\ 
 & Asia \& Oceania & 157 (7.9\%) \\ 
 & South America & 98 (5\%) \\ 
 & Africa \& Middle East & 36 (1.8\%) \\ 
 & Other \& Global & 125 (6.3\%) \\ \bottomrule
\end{tabularx}
\end{table}

\begin{table}[!ht]
\centering
\caption{Model Fit Indices}
\label{tab:modelfitindices}
\begin{tabular}{@{}m{5cm}m{2cm}m{6cm}@{}}
\toprule
\textbf{Model fit index}                                & \textbf{Value}      & \textbf{Interpretation}                                                                                                                                             \\ \midrule
Chi-Square ( $\chi^2$)                                            & 2,757.805            & n/a                                                                                                                                                                 \\ 
Degrees of freedom (df)                                 & 710                 & n/a                                                                                                                                                                 \\ 
CMIN/df                                                 & 3.884               & A value below 5 indicates an acceptable model fit~\cite{marsh1985application}, below 3 a good fit~\cite{hu1999cutoff}
\\ 
Root Mean Square Error of Approximation & .038 & Values $\leq .05$ indicates good model fit~\cite{byrne2010structural}  \\                                RMSEA 90\% CI   & .037-.040                                                                             \\ 
$p$ of Close Fit (PCLOSE)                                 & 1.000               & Probability that RMSEA $\leq 0.05$, where higher is better                                                                                                    \\ 
Comparative Fit Index (CFI)                             & .959               & Values $\geq .95$ indicates good model fit~\cite{hair2019multivariate}                                                                                                               \\ 
Tucker Lewis Index (TLI)                                & .953               & Values $\geq .92$ indicates good model fit~\cite{hair2019multivariate}                                                                                                                \\ 
Standardized Root Mean Square Residual (SRMR)           & .035              & Values $ \leq .05$ indicates good model fit~\cite{hair2019multivariate}                                                                                                                    \\ 
Team effectiveness: Variance explained by predictors ($R^2$)           & 75.6\%              & Values $ \geq 26\%$ indicates large effect~\cite{cohen1992power}                               
                                                \\ \addlinespace
Stakeholder satisfaction: Variance explained by predictors ($R^2$)           & 57.9\%              & Values $ \geq 26\%$ indicates large effect~\cite{cohen1992power}                               
                                                \\ \addlinespace
Team morale: variance explained by predictors ($R^2$)           & 34.9\%              & Values $ \geq 26\%$ indicates large effect~\cite{cohen1992power}                                       
\\ \bottomrule
\end{tabular}
\end{table}

\subsection{Hypothesis Testing and Interpretation}

\begin{table*}[!ht]
\centering
\small
\caption{Parameter Estimates, Confidence Intervals, Standard Errors, and Standardized Coefficients for Direct Effects and Indirect effects for hypotheses (statistically significant hypotheses at $p <0.05$ are set in boldface)}
\label{tab:parameterestimates}
\begin{tabularx}{\textwidth}{@{}Xlllll@{}}
\toprule
\textbf{Parameter} & \textbf{Unstandardized} & \textbf{95\% CI} & \textbf{SE} & \textit{\textbf{p}} & \textbf{Standardized} \\ \midrule
\multicolumn{6}{c}{\emph{Direct Effects}} \\ \addlinespace   
\textbf{H1: Responsiveness   $\rightarrow$ Team Effectiveness} & \textbf{.235} & \textbf{(.111, .418)} & \textbf{.050} & \textbf{.007} & \textbf{.402} \\
\textbf{H2: Stakeholder   Concern $\rightarrow$ Responsiveness} & \textbf{.400} & \textbf{(.265, .544)} & \textbf{.067} & \textbf{.002} & \textbf{.414} \\
\textbf{H3a: Stakeholder Concern   $\rightarrow$ Team Effectiveness} & \textbf{.291} & \textbf{(.138, .417)} & \textbf{.047} & \textbf{.015} & \textbf{.513} \\
\textbf{H4a: Continuous   Improvement $\rightarrow$ Stakeholder Concern} & \textbf{.616} & \textbf{(.491, .749)} & \textbf{.061} & \textbf{.001} & \textbf{.545} \\
\textbf{H4b: Continuous   Improvement $\rightarrow$ Responsiveness} & \textbf{.372} & \textbf{(.166, .582)} & \textbf{.082} & \textbf{.008} & \textbf{.340} \\
\textbf{H5a: Team Autonomy   $\rightarrow$ Continuous Improvement} & \textbf{.710} & \textbf{(.61, .825)} & \textbf{.054} & \textbf{.001} & \textbf{.655} \\
H5b: Team Autonomy $\rightarrow$   Stakeholder Concern & .135 & (.011, .273) & .061 & .074 & .110 \\
\textbf{H5c: Team Autonomy   $\rightarrow$ Responsiveness} & \textbf{.306} & \textbf{(.110, .513)} & \textbf{.081} & \textbf{.015} & \textbf{.258} \\
\textbf{H6a: Management Support   $\rightarrow$ Team Autonomy} & \textbf{.251} & \textbf{(.208, .292)} & \textbf{.020} & \textbf{.001} & \textbf{.427} \\
\textbf{H6b: Management Support   $\rightarrow$ Continuous Improvement} & \textbf{.107} & \textbf{(.067, .147)} & \textbf{.020} & \textbf{.001} & \textbf{.168} \\
\textbf{H6c: Management Support   $\rightarrow$ Stakeholder Concern} & \textbf{.184} & \textbf{(.143, .226)} & \textbf{.021} & \textbf{.001} & \textbf{.255} \\
H6d: Management Support $\rightarrow$ Responsiveness & -.027 & (-.078, .029) & .028 & .466 & -.039 \\
\multicolumn{6}{c}{\emph{Indirect Effects}} \\ \addlinespace
\textbf{H3b: Stakeholder Concern  $\rightarrow$ Responsiveness $\rightarrow$ Team Effectiveness} & \textbf{.094} & \textbf{(.044, .225)} & \textbf{.138} & \textbf{.004} & \textbf{.166} \\
\textbf{Continuous Improvement   $\rightarrow$ Responsiveness $\rightarrow$ Team Effectiveness} & \textbf{.087} & \textbf{(.047, .152)} & \textbf{.092} & \textbf{.004} & \textbf{.137} \\
\textbf{Team Autonomy $\rightarrow$   Responsiveness $\rightarrow$ Team Effectiveness} & \textbf{.072} & \textbf{(.015, .175)} & \textbf{.051} & \textbf{.016} & \textbf{.104} \\
\textbf{Team Autonomy   $\rightarrow$ Continuous Improvement $\rightarrow$ Responsiveness $\rightarrow$ Team   Effectiveness} & \textbf{.062} & \textbf{(.034, .108)} & \textbf{.023} & \textbf{.003} & \textbf{.223}
 \\ \bottomrule
\end{tabularx}
\end{table*}

We reported the means, standard deviations, and (Pearson) correlations of all variables in Table~\ref{tab:meansdeviationscorrelations} in the Appendix. Following recommendations in statistical literature~\cite{byrne2010structural,kline2015principles}, we used a bootstrapping procedure with 2,000 samples and 95\% bias-corrected confidence intervals to more accurately estimate parameters and their \textit{p}-values for direct effects, factor loadings, and the hypothesized indirect effects. This resulted in a standardized, bias-corrected estimate ($\beta$) for each path, along with a \textit{p}-value to test the null hypothesis of there being no effect at all. The parameter estimates for our hypotheses are reported in Table~\ref{tab:parameterestimates}, with additional estimates and factor loadings available in Table~\ref{tab:parameterestimatesextended} in the Appendix.

The results allowed us to confirm ten out of twelve hypotheses. 
We found a significant and positive association between teams' responsiveness and team effectiveness (H1, $\beta=.402,p<.01$). So more responsive teams are also more effective in terms of their team morale and stakeholder satisfaction. Regarding stakeholder concern, we found a direct positive effect on responsiveness (H2, $\beta=.414,p<.01$). This supports the notion that a high concern for stakeholder needs drives teams to be more responsive.
We also hypothesized that stakeholder concern influences team effectiveness, both directly (H3a) and mediated through responsiveness (H3b). This means that a high concern for stakeholder needs is not in itself beneficial when a team is not also responsive. The data supported partial mediation but not full mediation. Full mediation exists when there is only an indirect significant effect between two variables but not a direct one~\cite{baron1986moderator}. We did find a significant direct effect between stakeholder concern and team effectiveness (H3a, $\beta=.513,p=.02$) and also a significant indirect effect from stakeholder concern to team effectiveness (H3b, $\beta=.166,p<.01$). So while stakeholder concern in itself is a solid contributor to Scrum team effectiveness, the most effective Scrum teams are those that are pairing it with high responsiveness.

With responsiveness and stakeholder concern as the central contributors to team effectiveness, we hypothesized three team-level antecedents that create the right conditions. First, we hypothesized that teams are more likely to become responsive and develop a concern for stakeholders when they operate in a climate that encourages continuous improvement. We found support for this through a significant effect from continuous improvement to stakeholder concern (H4a, $\beta=.545,p<.01$) and (H4b, $\beta=.340,p<.01$). 
In a similar vein, we found support for the positive influence of team autonomy. A positive effect was found on continuous improvement (H5a, $\beta=.655,p<.01$) and on responsiveness (H5c, $\beta=.258,p=.02$), but not on stakeholder concern (H5b, $\beta=.110,p=.074$). So the more autonomous teams are, the more likely they are to improve continuously and the more responsive they can be.

Finally, we hypothesized that support from management is an important enabler for the other factors in our model. And while we found a positive effect from management support on team autonomy (H5a, $\beta=.427,p<.01$), continuous improvement (H5b, $\beta=.168,p<.01$) and stakeholder concern (H5c, $\beta=.255,p<.01$), management support was not significantly associated with responsiveness (H5d, $\beta=-.039,p=.466$).

\section{Discussion}
\label{sec:Discussion}
Teams are at the heart of the Scrum framework. To date, few studies have investigated how team-level factors contribute to Scrum team effectiveness. Moe et al. ~\cite{moe2008understanding,moe2010teamwork} were among the first to do so. They applied a  teamwork model by Salas et al. ~\cite{salas2005there} to understand how cognitive and behavioral processes allow team members to collaborate effectively. Nevertheless, teamwork is one piece of the puzzle. Because teams are open systems, their effectiveness is constrained by the larger systems they are part of~\cite{decuyper2010grasping}. This is particularly relevant to boundary-crossing challenges typical to Scrum teams, such as their interactions with stakeholders, the capability to release frequently, management support, team autonomy, and a climate of continuous improvement. Thus, this study contributes to existing research by offering a systematic perspective on the team-level factors that contribute to Scrum team effectiveness and manifest within the team and on the boundaries with larger systems.

We defined team effectiveness both from the perspective of external stakeholders (e.g., customers, users, and others outside the team) and internal stakeholders (team members). We used self-reported measures from team members to operationalize both perspectives and the success factors we expect to influence team effectiveness. To understand which team-level factors contribute to Scrum team effectiveness and answer RQ$_1$, we induced a five-factor model for team effectiveness from 13 case studies. To answer RQ$_2$, we tested the generalization of our model with Covariance-Based Structural Equation Modeling and a large sample of 1,978 Scrum teams and 4,940 respondents. Our sample included teams of varying levels of experience with Scrum from different regions in the world and different types and sectors of organizations (Table~\ref{tab:samplecomposition}).

Our model demonstrated a good fit based on recommended fit indices with this sample. The five factors explained a substantial amount of variance in our two measures for team effectiveness: stakeholder satisfaction ($R^2=58\%$) and team morale ($R^2=35\%$). 
Although Scrum teams are embedded in larger organizational systems, this suggests that much of what makes Scrum teams effective can be designed and assessed at the team level. We discuss the findings and their implications below and summarize them in Table~\ref{tab:implications}.

First, \textbf{Scrum teams that release to stakeholders frequently are indeed more effective than teams that do not}, as evidenced by higher stakeholder satisfaction and higher team morale (H1). We defined this more broadly as ``responsiveness'' and included release frequency and refinement. This finding reflects the premise of Agile methodologies that it is easier to manage complexity and discover emergent needs through many small releases~\cite{AgileManifesto2001} than through incidental large releases. It also stresses the motivational effect of frequent releases on teams that we observed in the case studies and is consistent with the evidence on motivation by Kahn~\cite{kahn1990psychological}. 

Second, we found that \textbf{responsiveness to stakeholders alone is not enough}. Scrum teams negotiate much of their effectiveness in collaborating with stakeholders and understanding what is important to them. This ``stakeholder concern'' is indicated through frequent collaborations with stakeholders, the use of Sprint Reviews to gather feedback, shared goals, and a focus on what is valuable. Teams that score high in these areas develop a clearer sense of value and purpose than teams that do not. Our results suggest that such teams are more likely to develop the capabilities to be responsive to stakeholder needs (H2). We also found that a high concern for stakeholder needs contributes to team effectiveness directly and is mediated through responsiveness (H3a \& H3b). This means that \textbf{while Scrum teams are indeed more effective when they understand the needs of their stakeholders, their actual ability to respond quickly to those needs further strengthens that effect}. In support of earlier studies by Bass et al. ~\cite{bass2018empirical} and Judy \& Krummins-Beens ~\cite{judykruminsbeens2008}, we found that Product Owners are instrumental in shaping stakeholder concern. Product Owners can support Scrum teams by providing a vision and strategy to address stakeholder needs, using formal gatherings to gather feedback from stakeholders, and encouraging collaboration between the team and stakeholders.

Third, \textbf{we identified a climate of continuous improvement as a crucial team-level antecedent of responsiveness and stakeholder concern, and thus Scrum team effectiveness}. Continuous improvement is indicated by high psychological safety, a focus on quality, shared learning with other teams, Sprint Retrospectives that result in actionable improvements, and an environment that encourages learning. Teams that score high in this area are more likely to be responsive (H4a) and show higher concern for stakeholder needs (H4b), which in turn increases their effectiveness. We theorized that this is because such teams are better able to continuously learn how to overcome emerging barriers to responsiveness and stakeholder collaboration and work with the larger system to affect change. Although the cross-sectional nature of our data does not allow us to support such a causal mechanism conclusively, the results are certainly consistent with research on organizational learning~\cite{marsick2003demonstrating,marsick1999facilitating} and learning as part of Agile transformations~\cite{hoda2017becoming}. Another way to think of our conceptualization of continuous improvement is a team's ability to learn. This is also consistent with Strode, Dings{\o}yr \& Lindsorn~\cite{strode2022teamwork}, who identify learning as an important influence on effective teamwork in Agile teams. Learning in particular is a process that is most effective when it crosses system boundaries~\cite{kasl1997teams,decuyper2010grasping}. 

Fourth, \textbf{we identified team autonomy as an essential team-level antecedent of continuous improvement and stakeholder concern, and thus of Scrum team effectiveness}. Team autonomy is indicated by a high level of self-management and cross-functionality. Teams that score high in this area tend to operate in climates that are more conducive to continuous improvement (H5a) and are more responsive (H5b). If team autonomy is seen as a reduction in external and internal constraints a team has to navigate to become more effective, the results illustrate different types of constraints. For responsiveness, we found a modest positive effect on team autonomy. The more autonomy teams experience, the more likely they are to take ownership over improvements to their work as a team~\cite{junker2021Agile}. As teams expand their autonomy over their composition, work method, and work distribution, they also tend to be more responsive to emerging needs. We found a much stronger positive effect of team autonomy on continuous improvement, which is corroborated by other research~\cite{decuyper2010grasping}. However, contrary to our expectations, we did not find an effect of team autonomy on stakeholder concern (H5c). We expected that the mandate of a team to make decisions about their product is one type of constraint associated with team autonomy. We expected this mandate to influence the ability of teams to focus on stakeholder concerns. We found in the case studies that some Product Owners controlled scope and delivery dates and were able to adapt to emerging stakeholder needs. In contrast, other Product Owners had to follow established requirements and lacked such flexibility. 
However, no such effect was found. Team autonomy did not predict stakeholder concern in our analyses. This suggests that the autonomy of Scrum teams to make decisions about how to do their work (self-management) is mainly orthogonal to their autonomy to decide what is most valuable to their stakeholders (product mandate). We are not aware of studies that specifically explored this distinction, so this may be interesting to explore in future studies. However, this finding may also reflect a limitation of how we operationalized team autonomy in the survey (see Section \ref{sec:limitations}).

A practical implication of these findings is that \textbf{organizations that seek to improve Scrum team effectiveness do well to invest in their autonomy}, which is supported by other studies of Agile teams~\cite{donmez2013practice,moe2010teamwork}. Moe et. al.~\cite{moe2019team,moe2009overcoming} see this as particularly challenging in large-scale environments where the desire to increase team autonomy often competes with a desire to align teams and coordinate decision-making. A more effective strategy in such environments is to encourage teams to self-align by involving them in shared goal-setting and the development of a shared vision~\cite{moe2019team}. Team autonomy is also difficult in ``specialist cultures'' where redundancy of skills in teams is considered an inefficiency and thus discouraged, resulting in more dependencies and lower individual autonomy~\cite{moe2009overcoming}. So while our study emphasizes the need for increased team autonomy, the preferred interventions are dependent on the context of a team.

Fifth, \textbf{this study stresses the importance of clear support from management}. We found that teams that experience such support also have higher team autonomy, continuous improvement, and stakeholder concern (H6a, H5b \& H6c). The results did not support the hypothesized positive effect on responsiveness (H6d). This also seems to be a result of different types of constraints. Where stakeholder concern, continuous improvement, and team autonomy are more susceptible to team-level constraints, responsiveness may be more constrained by individual skills, technologies, or legacy systems. A practical implication of these results is that support from management is most relevant where it involves the removal of constraints in the immediate environment of Scrum teams, particularly where it concerns the autonomy of Scrum teams and the quality and mandate of Product Owners. This is an excellent example of what Cummings~\cite{cummings2005organization} calls ``boundary control''; one of the strategies that managers can use to support autonomous workgroups.

Thus, our resulting model identifies five factors contributing to Scrum team effectiveness that can be observed, measured, and influenced at the team level. The factors in our model also reflect the dynamic reality of Scrum teams as part of larger systems and how they continuously negotiate their effectiveness on their boundaries with regards to their autonomy, learning, stakeholder collaboration, and responsiveness. With this study, we provide an additional perspective on Scrum team effectiveness as compared to Moe et al. ~\cite{moe2008understanding,moe2010teamwork}. Although teamwork did not naturally emerge from our case studies as a separate factor, teamwork's components and coordinating mechanisms can certainly be recognized in variables such as psychological safety, shared goals, cross-functionality, and self-management. 

We did not specifically investigate the role of characteristics like team size, project size, skill level, organization size, and legacy software on these processes. Such moderators did not emerge from our qualitative phase. Nevertheless, the resulting statistical model can explain a large amount of variance in team effectiveness without explicitly considering such moderators. However, we recognize that the predictive power will probably be larger when such moderators are considered. For example, releasing frequently functions that have several dependencies on a legacy codebase may be more difficult and hamper responsiveness~\cite{bosch2011introducing}. Teams may experience lower team autonomy as project size increases due to increased dependencies. It may also be harder for them to interact directly with stakeholders~\cite{van2013Agile}. 

\hyphenation{experience}
\begin{table}[!ht]
\centering
\tiny
\caption{Summary of Findings and Implications}
\label{tab:implications}
\begin{tabular}{p{2cm}p{5cm}p{5cm}}
\toprule
 & \textbf{Findings} & \textbf{Implications} \\ \midrule
\textbf{Five Factor Theory for Scrum Team Effectiveness}
& From 13 case studies, we developed a theoretical model for Scrum teams from thirteen lower-order indicators grouped into five latent factors that predict Scrum team effectiveness. This model fit the data from a large and representative sample of Scrum teams well ($Chi^2(710)=2,757.805$; $TLI=.953$; $CFI=.959$; $RMSEA=.038$; $SRMR=.035$). Together, the five factors explain a substantial amount of variance in the indicators we used to operationalize team effectiveness; stakeholder satisfaction (57.9\%) and team morale (34.9\%).
& Design and assess Scrum teams with five team-level factors in mind: responsiveness, stakeholder concern, continuous improvement, team autonomy, and management support. Design the environment of Scrum teams to minimize constraints to these factors on the one hand and train and support them in the skills they need for each factor. \\ \addlinespace
\textbf{Responsiveness}
& Responsiveness is positively associated with team effectiveness ($\beta=.402,p<.01$).
& Support Scrum teams in their ability to be responsive. Implement technical tooling, increase automation, and train necessary skills (particularly \textit{refinement}). Invest in team autonomy, stakeholder concern, and management support to make the need for responsiveness more relevant to teams. \\ \addlinespace
\textbf{Stakeholder Concern}
& The concern that teams show for their stakeholders is positively associated with responsiveness ($\beta=.414,p<.01$). We also found that stakeholder concern contributes to team effectiveness directly ($\beta = .513, p < .05$), and also partially through responsiveness ($\beta = .166, p < .01$). So the positive effect of stakeholder concern is strongest when paired with high responsiveness.
& Product Owners can increase Stakeholder Concern by co-opting teams in product strategy formulation, goal setting, and collaboration with stakeholders. If Scrum teams are unable to release frequently in the first place, efforts must be undertaken to remove organizational constraints, increase automation and build technical skills. \\ \addlinespace
\textbf{Continuous Improvement}
& The degree to which teams engage in continuous improvement is positively associated with stakeholder concern  ($\beta = .545, p < .001$) and responsiveness ($\beta = .340, p < .01$).
& Scrum teams are advised to direct their continuous improvement process towards the five key factors identified in this study: responsiveness, stakeholder concern, team autonomy, management support, and continuous improvement. These factors are most likely to highlight constraints to team effectiveness stemming from internal or external factors to the team. In turn, organizations should broaden the autonomy of teams to encourage teams to take control over improvements. \\ \addlinespace
\textbf{Team Autonomy}
& Team Autonomy was positively associated with continuous improvement ($\beta = .655, p < .001$) and responsiveness ($\beta = .258, p < .05$). Contrary to our expectations, we did not find an effect on stakeholder concerns.
& Expand the autonomy of Scrum teams primarily in two areas. The first is internal to teams and concerns the degree to which its members are cross-functional. The second concerns constraints imposed by the organizational environment that limit control over tooling, team composition, choice of process, and Product Owners' mandate over their product. \\ \addlinespace
\textbf{Management support}
& Management Support was found to be positively associated with team autonomy ($\beta = .427, p < .001$), continuous improvement ($\beta = .168, p < .001$) and stakeholder concern ($\beta = .255, p < .001$), but no significant effect was found on responsiveness.
& Management can most effectively contribute to Scrum teams by increasing their autonomy in terms of self-management and product mandate. Train management in the skills needed to support rather than direct. \\ \addlinespace
 \\ \bottomrule
\end{tabular}
\end{table}

\subsection{Limitations}
\label{sec:limitations}
This is a Mixed Methods study. Following the recommendations of Russo et al.~\cite{Russo2018ISQ}, we discuss both qualitative and quantitative limitations. In particular, we combined the threats to credibility, transferability, dependability, and confirmability~\cite{guba1981criteria} of our multiple field studies, with internal, construct, external, and conclusion validity from our sample study~\cite{wohlin2012experimentation}.

\textbf{Credibility \& Internal}. 
The factors that we identified in the inductive research phase were derived from thirteen cases over five years. Triangulation was achieved through the use of multiple independent cases. For the deductive research part of our study, we employed several strategies to safeguard internal validity. One threat to public surveys is that samples are based on voluntary (non-probabilistic) responses and may be biased due to self-selection or non-completion. Therefore, we advertised the survey on industry platforms that are frequented by the target population. We also incentivized participation by offering teams a detailed score profile upon completion. Although we can not calculate the non-response rate without knowing how many people visited the survey, we did establish that 42\% of the participants that started the survey also completed it.
The cross-sectional nature of the data also does not allow conclusions about the causality of effects.

\textbf{Transferability \& External}.
A key strength of the Eisenhardt approach is that it grounds theories in the substantive area of use~\cite{eisenhardt2007theory}, and thus increases the likelihood of valid and relevant theories~\cite{age2011grounded}. More importantly, the second phase of our study provided support for our theory with one of the most extensive sample studies ever performed in this area, surveying over 5,000 professionals of 2,000 Scrum teams from a range of industries, organizations, and countries (cf. Table~\ref{tab:samplecomposition}), leading to generalizable results. Although we used a non-random sample for the case studies, it included Scrum teams that varied in size, sector, scale, and product.
However, we can not make conclusions for teams that apply Scrum to other domains than software engineering or teams in general.

The data collection was partially performed during the COVID-19 pandemic. Some of the scales, like ``Team Morale'' and ``Psychological Safety'', may have been scored differently because teams worked from home. We added a 2-item 7-point Likert scale for a subset of teams ($N=351$) to assess how they expected their results to have been different outside of COVID-19. The team mean score ($M=2.81, SD = 1.63$) did not suggest a substantial impact. We also added the COVID-19 indicators as controls in our Structural Equation Model to further assess any confounding influence of COVID-19 on our results but found no significant effect on any of the other indicators in the survey.

\textbf{Dependability \& Construct}. 
Triangulation was achieved through multiple cases for five years.
Nevertheless, we can not exhaustively rule out personal bias in the case studies. 
Thus, we used well-established methodological guidelines to explain the research process and report our findings.
Also, the case studies were performed to inform theory, so the emphasis lies on the second phase of our study. 
To maximize content validity for the second phase, we used existing validated scales where available. Where this was not the case, we derived scales from observations that recurred across cases. As a result, the proposed measurement model fits the data well. A heterotrait-monotrait (HTMT) analysis confirmed discriminant validity for all measures. The reliability for all but one measure exceeded the cutoff recommended in the literature ($CR>=.70$~\cite{hair2019multivariate}).

Our survey consisted of self-reported items. We ran three pilots with the survey to identify problematic items, improve reliability and increase the response rate.
A confirmatory factor analysis showed that all items loaded primarily on their intended scales (see Table~\ref{tab:cfa}). The scales also demonstrated good overall reliability.
Although we statistically controlled for social desirability and common method bias, other confounding variables may have influenced how participants interpreted the items. This is particularly relevant to the operationalization of team effectiveness, which is based on self-reported scores on team morale and the perceived satisfaction of stakeholders. Mathieu et al. ~\cite{mathieu2008team} recognize that such affect-based measures may suffer from a ``halo effect''. Future studies could ask stakeholders to rate their satisfaction with team outcomes directly. This does not rule out a halo effect entirely but is conceptually closer to what matters to organizations. Future studies could also attempt to find more objective measures for team effectiveness.

Furthermore, our operationalization for team autonomy in the survey did not include a measure for product mandate. This may explain why we did not find the hypothesized effect of team autonomy on stakeholder concern. Instead, we may have indirectly measured such mandate through the indicators for stakeholder concern, more specifically value focus (``The Product Backlog of this team is ordered with a strategy in mind'', and ``The Product Owner of this team has a clear vision for the product.''). 

Our sample composition (Table~\ref{tab:samplecomposition}) shows that a wide range of teams participated in the questionnaire, with different levels of experience from different parts of the world and different types of organizations. We also observed a broad range of scores on the various measures. This provides confidence that a wide range of Scrum teams participated. Furthermore, our sample size and the aggregation of individual-level responses to team-level aggregates reduce variability due to non-systematic individual bias.

\textbf{Confirmability \& Statistical conclusion}. 
In this study, we used observations from thirteen exploratory case studies to inform a theoretical model about factors that contribute to the effectiveness of Scrum teams. In the second phase, we used Structural Equation Modeling to test the entire model simultaneously~\cite{kline2015principles, byrne2010structural}. The resulting model fits the data well on all fit indices recommended by statistical literature and explains a substantial amount of variance in the dependent variables. Finally, we published the team-level data and syntax files on Zenodo for reproducibility.

\section{Conclusion}
\label{sec:Conclusion}
This paper proposes a theoretical model to understand which team-level factors determine Scrum team effectiveness. To date, only a few studies have investigated this question (e.g,~\cite{moe2008understanding,strode2022teamwork}). This is surprising due to the prevalence of Scrum teams~\cite{version1stateofAgile}. Furthermore, Scrum teams have been identified as the most important success factor for Agile projects in the Agile Success Model~\cite{russo2021ASM}. This study contributes to existing research by offering a systematic perspective on the team-level factors that determine Scrum team effectiveness. A generalizable theory for Scrum teams offers practical guidance to organizations and practitioners on designing, assessing, and supporting teams.

We applied a Mixed Methods approach by first collecting qualitative data from thirteen case studies to develop a preliminary theoretical model, which we then subjected to a quantitative test with a large sample of Scrum teams ($N=1,978$). The data identified five high-order factors (responsiveness, stakeholder concern, continuous improvement, team autonomy, and management support) and thirteen lower-order factors (e.g., value focus, Sprint Review quality, cross-functionality, and refinement). Based on patterns in the observational data and supported by extant literature, we proposed six hypotheses to explain how these factors interact to make Scrum teams more or less effective, as indicated by their ability to satisfy stakeholders on the one hand and experience high team morale on the other.

Our findings highlight the interplay between stakeholder concern and responsiveness as drivers of Scrum team effectiveness. \textbf{The most effective teams are those that pair the ability to frequently release \textit{with} a strong focus on the needs of their stakeholders. In turn, this requires a high degree of team autonomy, continuous improvement, and support from management}.

The model in this paper acts as a grounding framework to inform future research on Scrum team effectiveness. We recognize several areas that are useful to explore in future studies. First, we proposed our model as a systemic alternative to the teamwork model by Moe et al. ~\cite{moe2008understanding}. It would be interesting to explore how both conceptualizations strengthen each other. 
Second, future studies can investigate how moderators such as project size, legacy systems, and skill-level of team members improve the predictive power of the Scrum Team Effectiveness model. Third, it is worthwhile to test the proposed causality with longitudinal or quasi-experimental designs. First, this may shed more light on which types of interventions are most relevant to practitioners and in which order. 
Fourth, it would be helpful to replicate our findings by directly measuring stakeholder satisfaction rather than indirectly through team members.
Finally, we expect that much of our model can be generalized to other types of Agile teams. Only a few factors and a few items in our questionnaire are specific to the Scrum framework (e.g., Sprint Review Quality). It would be helpful to test the model with a version of the questionnaire that is agnostic to the Agile methodology in use by a team.

\section{Acknowledgments}
\label{sec:Ack}
The authors would like to thank all the organizations and informants for their efforts and time. We would also like to thank Johannes Schartau for acting as an additional rater in the qualitative case studies.

\section{Supplementary Materials}
\label{sec:Supplement}
A replication package for the sample study is available at the following DOI to support Open Science: 10.5281/zenodo.5659120 under a CC-BY-NC-SA 4.0 license. The package includes the model definitions (AMOS), syntaxes for SPSS, and a fully anonymized, cleaned, and aggregated dataset of the analyzed Scrum teams. 

\section{Responsible disclosure}
\label{sec:Disclosure}
Christiaan Verwijs has a financial interest in The Liberators BV.

\bibliographystyle{ACM-Reference-Format}
\bibliography{bib}

\newpage
\begin{appendix}
\section{Appendix}
\label{sec:appendix}

\Small
\begin{longtable}{@{}p{.50\textwidth}ll@{}}
\hline
\textbf{1st Order Observations \& concepts} & \textbf{2nd Order Themes} & \textbf{Aggregate Dimension}\\ \hline
\endhead 
Stakeholders are (more) satisfied with the work done by the team & Stakeholder Satisfaction & Team Effectiveness \\
Stakeholders frequently complain about the quality of the product & & \\
Stakeholders give compliments to the team for a job well done& & \\
Team members feel motivated in their work& Team Morale & \\
Team members frequently complain, but there is no talk about how to solve things & & \\
Team members feel like they're in a hamster wheel (different day, same problems) & & \\
The team feels that iterations are artificial (i.e, "There's always a next Sprint" or "Lets just carry it over to the next Sprint") & Refinement & Responsiveness \\
Team members spend time during a Sprint to prepare work for the coming Sprints& &  \\
Sprint Planning tends to drag on as teams try to discover what they need to create that Sprint & &  \\
The Sprint Backlog contains only a handful of large Product Backlog Items &  & \\
Sprints generally result in at least one releasable increment for this team& Release Frequency & \\
Acceptance testing is done after the sprint, and not during & & \\
Sprint Reviews are used to review an Increment that is running on an environment that is as close to production as  possible (e.g, not a developers pc)& & \\
Sprint Reviews don't actually demonstrate working software (e.g, instead PowerPoint presentations are used, or the team simply talks about it) & & \\
The team frequently spends time cleaning up the Product Backlog by removing items that don't make sense anymore (e.g, in light of changed product vision, strategy)& Value Focus & Stakeholder Concern \\
Team members other than the Product Owner can reiterate the strategy behind the upcoming sprints& & \\
The team talks about how or why items on the Product Backlog are valuable to their stakeholders during various Scrum Events& & \\
The Backlog contains mostly technical tasks, and is difficult to connect to specific business needs by stakeholders & & \\
The team is familiar with the vision and strategy for the product they are working on& & \\
The work that a team does is aligned with other aspects of the business (support, sales, marketing, etc.)& Stakeholder Collaboration & \\
Except for the Product Owners, the members of a team don't generally speak with users, customers or other stakeholders & & \\
Product Owners involve customers/user/stakeholders in the prioritization of the backlog& & \\
Product Owners frequently meet with the customers or users using their software& & \\
The Backlog is not easily available to stakeholders, end-users and other teams & & \\
There is no interaction during Sprint Reviews between members of the team; people sit back and watch & & \\
Stakeholders have opportunities during Sprint Reviews to interact and review a working product or working features& Sprint Review Quality & \\
The team prepares for the Sprint Review by determining what they'd like feedback on and in what order& & \\
The Sprint Review is attended by users, customers and other stakeholders& & \\
The Product Owner uses the Sprint Review to get feedback from stakeholders on the work that was done& & \\
Most Sprints have a Sprint Goal that explains how the work in a Sprint ties together& Shared Goals & \\
There is a roadmap of Sprint Goals for the upcoming sprints& & \\
The Scrum Master is distributing work and chairing the Planning and the Sprint Review & Self Management & Team Autonomy \\
The Product Owner has a mandate to make decisions about the work that the team spends time on& & \\
The team is responsible for planning its own work capacity& & \\
Teams are involved in, or have complete control over, decisions that affect their composition or work-method& & \\
Teams are slowed down by dependencies on external teams or individuals who need to perform tasks that they can't do themselves & & \\
Teams are given responsibility only for specific functional areas (e.g, development, testing, design) & & \\
Individual team members are made responsible (and rewarded) for specific specializations & Cross-Functionality & \\
Team members collaborate on tasks from the Sprint Backlog, regardless of their specializations& & \\
Team members believe they are solely responsible for their specialization (e.g, "I'm only here to code") & &  \\
Team members actively keep each other up to date about what they are working on throughout the day& & \\
Teams are organized in terms of functional specializations (e.g, a 'continuous delivery team' or 'a design team') & &  \\
The testing for a particular item on the Sprint Backlog is performed by someone other than the developer (in or outside the team) & & \\
Sprint Retrospectives always yield the same issues over extended periods of time & Sprint Retrospective Quality & Continuous Improvement \\
Sprint Retrospectives are used to dive into deeper problems, not just the superficial symptoms& & \\
The action items from Sprint Retrospectives are usually not implemented & & \\
A lot of bugs and issues are discovered during the Sprint Review, or quickly after & Concern for Quality  &\\
Team members often talk about how to improve the quality of their work (e.g, during Sprint Retrospectives)& & \\
The Definition of Done doesn't define what high-quality looks like for a team & & \\
Team members are quick to shoot down ideas (e.g, "We've tried that before...", "That doesn't work   here") & Psychological Safety & \\
Team members generally offer vague and unclear contributions to the Daily Scrum (e.g, "I've was busy yesterday, I will be busy today") & & \\
Team members make an effort to listen to each other before sharing their own view& & \\
Team members openly ask for help when they are stuck& & \\
When a mistake is made, people are not singled out or punished for it& & \\
Teams share their learnings and challenges with other teams in the organization& Shared Learning & \\
Teams involve people from outside the team to work together on overcoming challenges they face& & \\
Team members express a desire to try new things or experiment with new technologies& Learning Environment &\\
Team members are motivated or encouraged to learn new things, read professional books, visit conferences or meetups& & \\
Management predominantly tells team members how to perform their work, rather than asking where they can support them& Management Support & Management Support \\
Management understands why this team works with Scrum& & \\
Management is available to help teams remove impediments& & \\ \hline
\caption{Data structure of observations and themes for the aggregate dimensions derived from case studies}
\label{tab:datastructure}
\end{longtable}

\pagebreak

\begin{table}[]
\tiny
\centering
\caption{Results of Structural Equation Modeling. Fit indices of the model and
standardized maximum likelihood estimates}
\label{tab:modelfitting}
\begin{tabularx}{\textwidth}{@{}llllllllX@{}}
\toprule
\textbf{Model} & \textbf{Chi-Square} & \textbf{Df} & \textbf{GFI} & \textbf{CFI} & \textbf{RMSEA} & \textbf{SRMR} & \textbf{TLI} & \textbf{Notes} \\ \midrule 
\multicolumn{8}{c}{\emph{Measurement Models}} \\ \addlinespace
MM1 & 3,307.80 & 898 & .93 & .957 & .037 & .038 & .949 & Base measurement model with only indicators and first-order latent   factors (covaried) \\
MM2 & 3,169.55 & 895 & .933 & .96 & .036 & .038 & .951 & Added covar between PS2 and PS3. PS1 and PS4. VF2 and VF3 due to high MI   (similar wording) \\
MM3 & 2,564.08 & 806 & .944 & .967 & .033 & .036 & .959 & Removed SH1 and PS5 due to many residual covariances with other items \\
MM4 & 2,324.47 & 763 & .934 & .97 & .032 & .028 & .963 & Removed RE1 due to low factor loading \\ \addlinespace
\multicolumn{8}{c}{\emph{Common Method Bias testing}} \\ \addlinespace
CMB1 & 1,946.49 & 720 & - & .977 & .029 & .021 & - & CMB with only a CLF (note: covariance matrix is not positive definite. so   a marker variable must be used instead) \\
CMB2 & 2,231.15 & 739 & - & .971 & .032 & .026 & - & CMB factor loads directly on all indicators (unconstrained) \\
CMB3 & 2,589.89 & 823 & - & .967 & .033 & .041 & - & CMB factor loads directly on all indicators (zero-constrained) \\
CMB4 & 2,524.85 & 822 & - & .968 & .032 & .033 & - & CMD factor loads directly on all indicators (constrained to same value to   test even distribution of bias) \\ \addlinespace
\multicolumn{8}{c}{\emph{Path Models}} \\ \addlinespace
PM1 & 2,757.81 & 710 & .932 & .959 & .038 & .035 & .952 & Base path model \\
\bottomrule
\end{tabularx}
\end{table}

\begin{table*}[]
\tiny
\centering
\caption{Scrum Team Survey Questionnaire}
\label{tab:survey}
\begin{tabularx}{\textwidth}{@{}llXl@{}}
\toprule
\textbf{Aggregate Dimension} & \textbf{Scale} & \textbf{Question} & \textbf{Type}\\ \midrule
 & Team & Where are the people that this team works  for - like users and customers - mostly based? & Single choice \\
 & Team & How many members does this team typically have, including a Scrum Master and Product Owner? & Single choice \\
 & Team & I consider this team to be very experienced with Scrum. & Likert (1-7) \\
 & Organisation & Which sector is this organisation mostly active in? & Single choice \\
 & Organisation Size & What is the size of this organisation? & Single choice \\
Responsiveness & Refinement & The Sprint Backlog of this team usually contains many small items. & Likert (1-7) \\
 & & During the Sprint, this team spends time to clarify work for the next   couple of Sprints. & Likert (1-7) \\
 & & During the Sprint, this team spends time breaking down work for coming   Sprints. & Likert (1-7) \\
 & Release Frequency & The majority of the Sprints of this team result in software that can be   deployed to production. & Likert (1-7) \\
 &  & For this team, most of the Sprints result in an increment that can be   released to users. & Likert (1-7) \\
Stakeholder Concern & Stakeholder Collaboration & Members of this team frequently meet with users or customers of what this team creates. & Likert (1-7) \\
 &  & People from this team often invite or visit people that use what this   team works on. & Likert (1-7) \\
 &  & People in this team closely collaborate with users, customers and other   stakeholders. & Likert (1-7) \\
 & Shared Goals & This team generally has clear Sprint Goals. & Likert (1-7) \\
 &  & During Sprint Planning, this team formulates a clear goal for the Sprint. & Likert (1-7) \\
 & Sprint Review Quality & The Product Owner of this team uses the Sprint Review to collect feedback   from stakeholders. & Likert (1-7) \\
 &  & During Sprint Reviews, stakeholders frequently try out what this team has been working on during the Sprint. & Likert (1-7) \\
 & Value Focus & The Product Owner of this team has a clear vision for the product. & Likert (1-7) \\
 &  & The Product Backlog of this team is ordered with a strategy in mind. & Likert (1-7) \\
 &  & Everyone in this team is familiar with the vision for the product. & Likert (1-7) \\
Continuous Improvement & Shared Learning & This team frequently works with other groups or teams to solve shared   problems. & Likert (1-7) \\
 &  & Teams in this organization share what they learn with other teams. & Likert (1-7) \\
 &  & Members from this team frequently meet with other teams to identify   improvements. & Likert (1-7) \\
 & Learning Environment & In and around this team, people are given time to support learning. & Likert (1-7) \\
 &  & In and around this team, people are rewarded for learning. & Likert (1-7) \\
 & Psychological Safety & In and around this team, people give open and honest feedback to each   other. & Likert (1-7) \\
 &  & In and around this team, people listen to the others' views before   speaking. & Likert (1-7) \\
 &  & In and around this team, whenever people state their view, they also ask   what others think.. & Likert (1-7) \\
 &  & In and around this team, people openly discuss mistakes in order to learn   from them. & Likert (1-7) \\
 &  & In and around this team, people help each other learn. & Likert (1-7) \\
 & Quality & Members of this team have a shared understanding of what quality means to   them. & Likert (1-7) \\
 &  & People in this team frequently talk about quality and how to improve it. & Likert (1-7) \\
 & Sprint Retrospective Quality & The Sprint Retrospectives of this team generally result in at least one   useful improvement. & Likert (1-7) \\
 &  & During Sprint Retrospectives, this team openly talks about improvements. & Likert (1-7) \\
 &  & This team uses Sprint Retrospectives to explore solutions for persistent challenges. & Likert (1-7) \\
Team Autonomy & Cross-Functionality & Most people in this team have the ability to solve the problems that come   up in their work. & Likert (1-7) \\
 &  & Everyone in this team has more than enough training and experience for the kind of work they have to do. & Likert (1-7) \\
 & Self-Management & This team has control over the scheduling of teamwork. & Likert (1-7) \\
 &  & This team is free to choose the method(s) to use in carrying out work. & Likert (1-7) \\
 &  & This team is able to choose the way to go about its work. & Likert (1-7) \\
Management Support & Management Support & People in a management position generally understand why this team works with Scrum. & Likert (1-7) \\
 &  & People in a management position help this team work with Scrum. & Likert (1-7) \\
Team Effectiveness & Stakeholder Satisfaction & Stakeholders frequently compliment this team with their work. & Likert (1-7) \\
 &  & Stakeholders are generally happy with the software this team delivers. & Likert (1-7) \\
 &  & Stakeholders are generally happy with how fast this team responds to   their needs. & Likert (1-7) \\
 &  & Our stakeholders compliment us with the value that we deliver to them. & Likert (1-7) \\
 & Team Morale & I am proud of the work that I do for this team. & Likert (1-7) \\
 &  & I am enthusiastic about the work that I do for this team. & Likert (1-7) \\
 &  & I find the work that I do for this team full of meaning and purpose. & Likert (1-7) \\
 Control Variables & Social Desirability & I am always courteous even to people who are disagreeable. & Likert (1-7) \\
 &  & No matter who I'm talking to, I'm always a good listener. & Likert (1-7) \\
 &  & I never feel resentful when I don't get my way. & Likert (1-7) \\
  & Covid 19 Control & The Covid19 pandemic, and working from home, has impacted my answers on   this survey & Likert (1-7) \\
 &  & I would have provided different answers if the pandemic never happened. & Likert (1-7) \\
\bottomrule
\end{tabularx}
\end{table*}

\begin{table}[]
\tiny
\centering
\caption{Results of Confirmatory Factor Analysis. Principal Components Analysis with Oblimin rotation and Kaiser normalization. Items and components are ordered by extraction. Factor loadings below .30 have been suppressed for readability.}.
\label{tab:cfa}
\resizebox{\textwidth}{!}{%
\begin{tabular}{@{}Xllllllllllllllllllll@{}}
\toprule
 \textbf{Item} & \textbf{1} & \textbf{2} & \textbf{3} & \textbf{4} & \textbf{5} & \textbf{6} & \textbf{7} & \textbf{8} & \textbf{9} & \textbf{10} & \textbf{11} & \textbf{12} & \textbf{13} & \textbf{14} & \textbf{15} & \textbf{16} & \textbf{17} \\ \midrule
PsychologicalSafety2 & .719 &  &  &  &  &  &  &  &  &  &  &  &  &  &  &  &  \\
PsychologicalSafety3 & .688 &  &  &  &  &  &  &  &  &  &  &  &  &  &  &  &  \\
PsychologicalSafety1 & .604 &  &  &  &  &  &  &  &  &  &  &  &  &  &  &  &  \\
PsychologicalSafety4 & .560 &  &  &  &  &  &  &  &  &  &  &  &  &  &  &  &  \\
StakeholderCollaboration1 &  & .943 &  &  &  &  &  &  &  &  &  &  &  &  &  &  &  \\
StakeholderCollaboration3 &  & .792 &  &  &  &  &  &  &  &  &  &  &  &  &  &  &  \\
StakeholderCollaboration2 &  & .790 &  &  &  &  &  &  &  &  &  &  &  &  &  &  &  \\
TeamMorale2 &  &  & .893 &  &  &  &  &  &  &  &  &  &  &  &  &  &  \\
TeamMorale2 &  &  & .865 &  &  &  &  &  &  &  &  &  &  &  &  &  &  \\
TeamMorale3 &  &  & .845 &  &  &  &  &  &  &  &  &  &  &  &  &  &  \\
SharedLearning1 &  &  &  & -.809 &  &  &  &  &  &  &  &  &  &  &  &  &  \\
SharedLearning3 &  &  &  & -.761 &  &  &  &  &  &  &  &  &  &  &  &  &  \\
SharedLearning2 &  &  &  & -.719 &  &  &  &  &  &  &  &  &  &  &  &  &  \\
ReleaseFrequency1 &  &  &  &  & .931 &  &  &  &  &  &  &  &  &  &  &  &  \\
ReleaseFrequency2 &  &  &  &  & .903 &  &  &  &  &  &  &  &  &  &  &  &  \\
SelfManagement2 &  &  &  &  &  & .868 &  &  &  &  &  &  &  &  &  &  &  \\
SelfManagement3 &  &  &  &  &  & .787 &  &  &  &  &  &  &  &  &  &  &  \\
SelfManagement1 &  &  &  &  &  & .664 &  &  &  &  &  &  &  &  &  &  &  \\
ManagementSupport2 &  &  &  &  &  &  & -.931 &  &  &  &  &  &  &  &  &  &  \\
ManagementSupport1 &  &  &  &  &  &  & -.926 &  &  &  &  &  &  &  &  &  &  \\
SocialDesirability2 &  &  &  &  &  &  &  & .863 &  &  &  &  &  &  &  &  &  \\
SocialDesirability1 &  &  &  &  &  &  &  & .854 &  &  &  &  &  &  &  &  &  \\
Refinement2 &  &  &  &  &  &  &  &  & -.917 &  &  &  &  &  &  &  &  \\
Refinement3 &  &  &  &  &  &  &  &  & -.893 &  &  &  &  &  &  &  &  \\
SharedGoals2 &  &  &  &  &  &  &  &  &  & -.972 &  &  &  &  &  &  &  \\
SharedGoals1 &  &  &  &  &  &  &  &  &  & -.953 &  &  &  &  &  &  &  \\
CrossFunctionality1 &  &  &  &  &  &  &  &  &  &  & -.867 &  &  &  &  &  &  \\
CrossFunctionality2 &  &  &  &  &  &  &  &  &  &  & -.772 &  &  &  &  &  &  \\
StakeholderHappiness2 &  &  &  &  &  &  &  &  &  &  &  & -.840 &  &  &  &  &  \\
StakeholderHappiness4 &  &  &  &  &  &  &  &  &  &  &  & -.836 &  &  &  &  &  \\
StakeholderHappiness3 &  &  &  &  &  &  &  &  &  &  &  & -.791 &  &  &  &  &  \\
SprintRetrospectiveQuality1 &  &  &  &  &  &  &  &  &  &  &  &  & .808 &  &  &  &  \\
SprintRetrospectiveQuality3 &  &  &  &  &  &  &  &  &  &  &  &  & .790 &  &  &  &  \\
SprintRetrospectiveQuality2 &  &  &  &  &  &  &  &  &  &  &  &  & .668 &  &  &  &  \\
ValueFocus1 &  &  &  &  &  &  &  &  &  &  &  &  &  & -.803 &  &  &  \\
ValueFocus2 &  &  &  &  &  &  &  &  &  &  &  &  &  & -.724 &  &  &  \\
ValueFocus3 &  &  &  &  &  &  &  &  &  &  &  &  &  & -.592 &  &  &  \\
SprintReviewQuality1 &  &  &  &  &  &  &  &  &  &  &  &  &  &  & .778 &  &  \\
SprintReviewQuality2 &  &  &  &  &  &  &  &  &  &  &  &  &  &  & .761 &  &  \\
LearningEnvironment2 &  &  &  &  &  &  &  &  &  &  &  &  &  &  &  & .818 &  \\
LearningEnvironment1 &  &  &  &  &  &  &  &  &  &  &  &  &  &  &  & .727 &  \\
ConcernForQuality2 &  &  &  &  &  &  &  &  &  &  &  &  &  &  &  &  & -.810 \\
ConcernForQuality1 &  &  &  &  &  &  &  &  &  &  &  &  &  &  &  &  & -.794
\\ \bottomrule
\end{tabular}
}
\end{table}

\begin{table}[]
\tiny
\centering
\caption{Means, Standard Deviations, and Correlations (Pearson) for first-order latent factors. All correlations are significant at $p < 0.001$}.
\label{tab:meansdeviationscorrelations}
\resizebox{\textwidth}{!}{%
\begin{tabular}{@{}Xllllllllllllllllllll@{}}
\toprule
 & \textbf{Factor} & \textbf{Mean} & \textbf{SD} & \textbf{1} & \textbf{2} & \textbf{3} & \textbf{4} & \textbf{5} & \textbf{6} & \textbf{7} & \textbf{8} & \textbf{9} & \textbf{10} & \textbf{11} & \textbf{12} & \textbf{13} & \textbf{14} & \textbf{15} & \textbf{16} \\ \midrule
1 & Learning Environment & 4.57 & 1.15 &  &  &  &  &  &  &  &  &  &  &  &  &  &  &  &  \\
2 & Concern for Quality & 4.52 & 1.12 & 0.76 &  &  &  &  &  &  &  &  &  &  &  &  &  &  &  \\
3 & Shared Learning & 3.62 & 0.99 & 0.75 & 0.79 &  &  &  &  &  &  &  &  &  &  &  &  &  &  \\
4 & Sprint Retrospective Quality & 4.93 & 1.22 & 0.73 & 0.75 & 0.63 &  &  &  &  &  &  &  &  &  &  &  &  &  \\
5 & Psychological Safety & 5.06 & 1.12 & 0.87 & 0.85 & 0.76 & 0.85 &  &  &  &  &  &  &  &  &  &  &  &  \\
6 & Shared Goals & 4.30 & 1.61 & 0.50 & 0.55 & 0.53 & 0.56 & 0.57 &  &  &  &  &  &  &  &  &  &  &  \\
7 & Stakeholder Collaboration & 3.74 & 1.35 & 0.53 & 0.50 & 0.59 & 0.35 & 0.50 & 0.37 &  &  &  &  &  &  &  &  &  &  \\
8 & Sprint Review Quality & 3.53 & 1.20 & 0.56 & 0.59 & 0.63 & 0.52 & 0.57 & 0.62 & 0.76 &  &  &  &  &  &  &  &  &  \\
9 & Value Focus & 4.76 & 1.26 & 0.68 & 0.71 & 0.66 & 0.64 & 0.70 & 0.64 & 0.59 & 0.74 &  &  &  &  &  &  &  &  \\
10 & Cross-Functionality & 4.57 & 0.88 & 0.69 & 0.77 & 0.67 & 0.66 & 0.79 & 0.52 & 0.49 & 0.52 & 0.66 &  &  &  &  &  &  &  \\
11 & Self-Management & 5.20 & 1.10 & 0.73 & 0.61 & 0.56 & 0.64 & 0.73 & 0.48 & 0.44 & 0.45 & 0.59 & 0.82 &  &  &  &  &  &  \\
12 & Refinement & 4.31 & 1.30 & 0.53 & 0.60 & 0.51 & 0.61 & 0.54 & 0.49 & 0.28 & 0.47 & 0.57 & 0.49 & 0.49 &  &  &  &  &  \\
13 & Release Frequency & 4.48 & 1.48 & 0.45 & 0.60 & 0.47 & 0.49 & 0.54 & 0.42 & 0.45 & 0.52 & 0.48 & 0.56 & 0.47 & 0.48 &  &  &  &  \\
14 & Team Morale & 5.12 & 1.08 & 0.60 & 0.64 & 0.61 & 0.61 & 0.69 & 0.47 & 0.46 & 0.52 & 0.65 & 0.68 & 0.63 & 0.40 & 0.45 &  &  &  \\
15 & Stakeholder Satisfaction & 4.71 & 1.05 & 0.62 & 0.67 & 0.62 & 0.60 & 0.67 & 0.47 & 0.61 & 0.68 & 0.70 & 0.70 & 0.61 & 0.47 & 0.61 & 0.71 &  &  \\
16 & Management Support & 4.63 & 1.29 & 0.68 & 0.58 & 0.59 & 0.50 & 0.59 & 0.44 & 0.45 & 0.58 & 0.66 & 0.57 & 0.60 & 0.44 & 0.41 & 0.56 & 0.60 &  \\
 & \textbf{Control variables} &  &  &  &  &  &  &  &  &  &  &  &  &  &  &  &  &  &  \\
17 & Social Desirability & 4.37 & 0.69 & 0.55 & 0.54 & 0.53 & 0.54 & 0.64 & 0.41 & 0.36 & 0.39 & 0.49 & 0.63 & 0.58 & 0.36 & 0.37 & 0.67 & 0.58 & 0.50
\\ \bottomrule
\end{tabular}
}
\end{table}

\begin{table*}[!ht]
\centering
\tiny
\caption{Parameter Estimates, Confidence Intervals, Standard Errors, Standardized Coefficients for Direct Effects and Indirect effects for hypotheses (statistically significant hypotheses at $p <0.05$ are set in boldface), and Factor Loadings}
\label{tab:parameterestimatesextended}
\begin{tabularx}{\textwidth}{@{}Xlllll@{}}
\toprule
\textbf{Parameter} & \textbf{Unstandardized} & \textbf{95\% CI} & \textbf{SE} & \textit{\textbf{p}} & \textbf{Standardized} \\ \midrule
\multicolumn{6}{c}{\emph{Direct Effects}} \\ \addlinespace   
\textbf{H1: Responsiveness   $\rightarrow$ Team Effectiveness} & \textbf{.235} & \textbf{(.111, .418)} & \textbf{.050} & \textbf{.007} & \textbf{.402} \\
\textbf{H2: Stakeholder   Concern $\rightarrow$ Responsiveness} & \textbf{.400} & \textbf{(.265, .544)} & \textbf{.067} & \textbf{.002} & \textbf{.414} \\
\textbf{H3: Stakeholder Concern   $\rightarrow$ Team Effectiveness} & \textbf{.291} & \textbf{(.138, .417)} & \textbf{.047} & \textbf{.015} & \textbf{.513} \\
\textbf{H4a: Continuous   Improvement $\rightarrow$ Stakeholder Concern} & \textbf{.616} & \textbf{(.491, .749)} & \textbf{.061} & \textbf{.001} & \textbf{.545} \\
\textbf{H4b: Continuous   Improvement $\rightarrow$ Responsiveness} & \textbf{.372} & \textbf{(.166, .582)} & \textbf{.082} & \textbf{.008} & \textbf{.340} \\
\textbf{H5a: Team Autonomy   $\rightarrow$ Continuous Improvement} & \textbf{.710} & \textbf{(.61, .825)} & \textbf{.054} & \textbf{.001} & \textbf{.655} \\
H5b: Team Autonomy $\rightarrow$   Stakeholder Concern & .135 & (.011, .273) & .061 & .074 & .110 \\
\textbf{H5c: Team Autonomy   $\rightarrow$ Responsiveness} & \textbf{.306} & \textbf{(.110, .513)} & \textbf{.081} & \textbf{.015} & \textbf{.258} \\
\textbf{H6a: Management Support   $\rightarrow$ Team Autonomy} & \textbf{.251} & \textbf{(.208, .292)} & \textbf{.020} & \textbf{.001} & \textbf{.427} \\
\textbf{H6b: Management Support   $\rightarrow$ Continuous Improvement} & \textbf{.107} & \textbf{(.067, .147)} & \textbf{.020} & \textbf{.001} & \textbf{.168} \\
\textbf{H6c: Management Support   $\rightarrow$ Stakeholder Concern} & \textbf{.184} & \textbf{(.143, .226)} & \textbf{.021} & \textbf{.001} & \textbf{.255} \\
H6d: Management Support $\rightarrow$ Responsiveness & -.027 & (-.078, .029) & .028 & .466 & -.039 \\
\multicolumn{6}{c}{\emph{Indirect Effects}} \\ \addlinespace
\textbf{H3: Stakeholder Concern   $\rightarrow$ Responsiveness $\rightarrow$ Team Effectiveness} & \textbf{.094} & \textbf{(.044, .225)} & \textbf{.138} & \textbf{.004} & \textbf{.166} \\
\textbf{Continuous Improvement   $\rightarrow$ Responsiveness $\rightarrow$ Team Effectiveness} & \textbf{.087} & \textbf{(.047, .152)} & \textbf{.092} & \textbf{.004} & \textbf{.137} \\
\textbf{Team Autonomy $\rightarrow$   Responsiveness $\rightarrow$ Team Effectiveness} & \textbf{.072} & \textbf{(.015, .175)} & \textbf{.051} & \textbf{.016} & \textbf{.104} \\
\textbf{Team Autonomy   $\rightarrow$ Continuous Improvement $\rightarrow$ Responsiveness $\rightarrow$ Team   Effectiveness} & \textbf{.062} & \textbf{(.034, .108)} & \textbf{.023} & \textbf{.003} & \textbf{.223} \\
\multicolumn{6}{c}{\emph{Factor Loadings}} \\ \addlinespace
Self-Management   $\rightarrow$ Team Autonomy &  & (.682, .789) &  &  & .739 \\
Cross-Functionality $\rightarrow$ Team Autonomy & .882 & (.736, .848) & .050 & .001 & .791 \\
Learning Environment $\rightarrow$ Continuous Improvement &  & (.706, .782) &  &  & .745 \\
Concern for Quality $\rightarrow$ Continuous Improvement & 1.016 & (.759, .825) & .052 & .001 & .792 \\
Shared Learning $\rightarrow$ Continuous Improvement & .786 & (.636, .709) & .046 & .001 & .672 \\
Sprint Retrospective Quality $\rightarrow$ Continuous Improvement & 1.044 & (.676, .744) & .054 & .001 & .713 \\
Psychological Safety $\rightarrow$ Continuous Improvement & 1.004 & (.824, .883) & .05 & .001 & .856 \\
Sprint Goals $\rightarrow$ Stakeholder Concern &  & (.531, .605) &  &  & .571 \\
Stakeholder Collaboration $\rightarrow$ Stakeholder Concern & .780 & (.517, .597) & .054 & .001 & .558 \\
Sprint Review Quality $\rightarrow$ Stakeholder Concern & 1.124 & (.701, .782) & .062 & .001 & .741 \\
Value Focus $\rightarrow$ Stakeholder Concern & 1.089 & (.784, .847) & .053 & .001 & .818 \\
Refinement $\rightarrow$ Responsiveness & .840 & (.516, .626) & .055 & .001 & .570 \\
Release Frequency $\rightarrow$  Responsiveness &  & (.525, .637) &  &  & .582 \\
Team Effectiveness $\rightarrow$  Stakeholder Happiness & 1.336 & (.715, .805) & .078 & .001 & .761 \\
Team Effectiveness $\rightarrow$ Team Morale &  & (.542, .635) &  &  & .591
 \\ \bottomrule
\end{tabularx}
\end{table*}

\end{appendix}

\end{document}